\documentclass[twocolumn]{aastex61}
\usepackage{amsmath}

\received{}
\revised{}
\accepted{}
\submitjournal{ApJ}

\shorttitle{Coronal density and temperature profiles}
\shortauthors{Pascoe et al.}

\begin{document}

\title{Coronal density and temperature profiles calculated by\\ forward modeling EUV emission observed by SDO/AIA}

\correspondingauthor{D. J. Pascoe}
\email{david.pascoe@kuleuven.be}

\author[0000-0002-0338-3962]{D. J. Pascoe}
\affiliation{Centre for mathematical Plasma Astrophysics, Mathematics Department, KU Leuven, Celestijnenlaan 200B bus 2400, B-3001 Leuven, Belgium}

\author[0000-0001-9500-6429]{A. Smyrli}
\affiliation{Jeremiah Horrocks Institute, University of Central Lancashire, PR1 2HE, UK}

\author[0000-0001-9628-4113]{T. Van Doorsselaere}
\affiliation{Centre for mathematical Plasma Astrophysics, Mathematics Department, KU Leuven, Celestijnenlaan 200B bus 2400, B-3001 Leuven, Belgium}

\begin{abstract}

We present a model for the intensity of optically thin EUV emission for a plasma atmosphere.
We apply our model to the solar corona as observed using the six optically thin EUV channels of the SDO/AIA instrument.
The emissivity of the plasma is calculated from the density and temperature using CHIANTI tables and the intensity is then determined by integration along the line of sight.
We consider several different profiles for the radial density and temperature profiles, each of which are constrained by the observational data alone with no further physical assumptions.
We demonstrate the method first by applying it to a quiet region of the corona, and then use it as the background component of a model including coronal holes, allowing the plasma densities and temperatures inside and outside the hole to be estimated.
We compare our results with differential emission measure (DEM) inversions.
More accurate estimates for the coronal density and temperature profiles have the potential to help constrain plasma properties such as the magnetic field strength when used in combination with methods such as seismology. 

\end{abstract}


\section{Introduction}
\label{sect:intro}

Extreme ultraviolet (EUV) imagers such as the Transition Region And Coronal Explorer \citep[TRACE; ][]{1999SoPh..187..229H} or the Atmospheric Imaging Assembly \citep[AIA; ][]{2012SoPh..275...17L} onboard the Solar Dynamics Observatory (SDO) have provided us a view of the solar corona free from the overwhelming intensity of white light produced by the much cooler and denser photosphere.
However, the interpretation of this data can be complicated by the optically thin nature of the emission which means the signatures of coronal structures or waves are integrated along the observational line of sight \citep[LOS; e.g.][]{2003A&A...397..765C,2008SoPh..252..101D,2012ApJ...746...31D}.

Under the approximation of a constant emissivity the intensity is determined by the square of the plasma density integrated along the LOS.
This approximation has been applied to study the transverse intensity profile of isothermal coronal loops \citep[e.g.][]{2003ApJ...598.1375A,2005ApJ...633..499A,2017A&A...600L...7P,2017A&A...605A..65G}.
This method was extended by \citet{2018ApJ...860...31P} to consider the emission from two overlapping loop legs and used simultaneously with seismological information from kink oscillations to infer the transverse density profile.
However, such methods are limited to investigating properties such as density contrast ratios or transverse scales rather than absolute values of the physical quantities.
Other examples of modeling coronal structures include the 3D reconstruction of streamers based on a slab model viewed from different angles either taken several days apart \citep{2006ApJ...642..523T} or by different instruments \citep{2018cosp...42E.796D}.

The FoMo code \citep{2016FrASS...3....4V} was developed to perform detailed forward modeling of EUV emission for coronal plasmas, and particularly for interpretation of results from numerical models or simulations.
For example, \citet{2016ApJS..223...23Y,2016ApJS..223...24Y} considered the appearance of standing kink oscillations of coronal loops in SDO/AIA images and further studies include additional effects such as the Kelvin-Helmholtz instability \citep[e.g.][]{2016ApJ...823...82M,2018A&A...610L...9K}. This allows the detailed observational signatures \citep[e.g.][]{2017ApJ...836..219A,2018ApJ...863..167G} to be predicted which is crucial for identification when these effects may be close to the limit of detectability with current instruments.
FoMo uses information from the CHIANTI atomic database \citep[e.g.][]{1997A&AS..125..149D,2009A&A...498..915D,2015A&A...582A..56D} to determine the emissivity of the plasma based on the density and temperature.

In addition to the EUV intensity, other observational signatures may be forward modeled such as Doppler shifts \citep[e.g.][]{2014SoPh..289.1959T,2017ApJ...836..219A} or gyrosynchrotron emission \citep[e.g.][]{2014ApJ...785...86R}.
The GX Simulator tool \citep[e.g][]{2011SPD....42.1811N,2015ApJ...799..236N} was developed to use information from magnetic extrapolations, radio, and X-ray observations to perform 3D modeling of solar structures and activity such as spatially-resolved flaring emission \citep{2015SoPh..290...79K}.
The FORWARD toolset \citep{2016FrASS...3....8G} can also be used to generate multiple observables for a prescribed physical state, including infrared, visible, EUV, and radio emission.

Coronal density and temperature profiles have previously been investigated by several observational campaigns.
\citet{1998ApJ...505..999F} used the Solar Ultraviolet Measurements of Emitted Radiation (SUMER) instrument onboard the Solar and Heliospheric Observatory (SOHO) spacecraft to study the plasma above the solar equator (quiet corona) and north pole (coronal hole), finding temperatures of $1.35$ and $0.83$~MK, respectively.
\citet{2002ApJ...571..999W} used SUMER to study a quiet coronal streamer and observed an increase in temperature above the limb ($r < 1.2 R_{\odot}$).
The simultaneous use of multiple observations to reconstruct the three-dimensional structure of the corona has also been considered by several authors.
In particular, the Solar Terrestrial Relations Observatory (STEREO) spacecraft provided two simultaneous viewpoints of the corona.
Differential emission measure (DEM) tomography \citep{2005ApJ...628.1070F} was applied to study the global corona \citep{2009ApJ...701..547F} and prominence cavities \citep{2009SoPh..256...73V}, including the calculation of maps of the estimated electron density.
STEREO has also been used for reconstruction of coronal loop geometry \citep[e.g.][]{2007ApJ...671L.205F,2008ApJ...679..827A}, estimating their density and temperature \citep{2008ApJ...680.1477A}, and used in conjunction with seismology \citep{2009ApJ...698..397V}.
Contact with STEREO-B has now been lost but it remains important to develop multi-instrument diagnostic methods.
At larger heights, coronal plasma has been studied using ground and space-based coronagraphs, such as the white light coronagraphs onboard the
Spartan 201-01 spacecraft \citep[e.g.][]{1995ApJ...447L.139F} and
Skylab \citep[e.g.][]{1977SoPh...55..121S,1996ApJ...458..817G}.
Here the radial density profile is often considered in terms of empirically derived polynomials \citep[e.g.][]{1950BAN....11..135V,1999ApJ...510L..63E}.

In this paper we present a method for estimating the density and temperature profiles of the solar corona by forward modeling the optically thin EUV emission and comparing the results with data from SDO/AIA.
Our models for the density and temperature profiles are described in Section~\ref{sect:profiles}.
We demonstrate our method by applying it to the quiet corona and coronal holes in Section~\ref{sect:fwdmod}.
Conclusions are presented in Section~\ref{sect:conclusions}.

\section{Coronal density and temperature profiles}
\label{sect:profiles}

In this section we describe our model for the coronal density and temperature profiles.
These profiles correspond to the background plasma i.e. excluding the presence of structures formed by the magnetic field such as coronal loops or holes.
We intend to obtain these profiles by forward modeling the corresponding EUV emission and comparing to observational data, without any assumed relationship between density and temperature.
However, it is instructive to first consider the behaviour expected from a simple hydrostatic equilibrium.

Ignoring the effect of a magnetic field, the hydrostatic equilibrium condition for a gas with pressure $P$ and mass density $\rho$ in with a gravitational field with acceleration $\mathbf{g}$ is

\begin{equation}
\nabla P = \rho \mathbf{g}.
\label{eq:hydrostatic}
\end{equation}
The ideal gas law further relates pressure and density to the temperature $T$ as $P = \rho R T / \tilde{\mu}$, where $R$ is the gas constant and $\tilde{\mu}$ is the mean molar mass.
Combining these gives

\begin{equation}
\frac{d P \left( r \right)}{d r} = - \rho \left( r \right) g \left( r \right) = - \frac{P \left( r \right)}{H \left( r \right)},
\label{eq:eqm}
\end{equation}
defining the pressure scale height $H \left( r \right) = R T \left( r \right) / \tilde{\mu} g \left( r \right)$.

First, we consider the case of the gravitational acceleration being constant
$\mathbf{g} = - g_{0} \hat{\mathbf{r}}$, where 
$g_0 = 274$~m s$^{-1}$ is the value at the nominal solar radius $R_{\odot} = 695.7$~Mm
and
$\hat{\mathbf{r}}$ is the unit vector in the radial direction.
For an isothermal atmosphere with temperature $T_{0}$, the scale height is constant and we obtain the density profile

\begin{equation}
\rho \left( r \right) = \rho_{0} \exp \left( \frac{R_{\odot} - r}{H_{0}} \right),
\label{eq:isothermal_gconst}
\end{equation}
where $\rho_{0} = \rho \left( r=R_{\odot} \right)$ and the isothermal scale height is $H_{0} = R T_{0} / \tilde{\mu} g_{0}$.

The assumption of a constant gravitational acceleration is valid for distances which are small in comparison to the radius of the Sun, $r \ll R_{\odot}$.
However, the size of the corona is comparable to $R_{\odot}$.
The gravitational acceleration decreases with height as $g \left( r \right) = g_{0} R_{\odot}^2/r^2$ and the corresponding density profile for an isothermal atmosphere is
\begin{equation}
\rho \left( r \right) = \rho_{0} \exp \left[ \frac{R_{\odot}}{H_{0}} \left( \frac{R_{\odot}}{r} - 1 \right) \right].
\label{eq:isothermal}
\end{equation}
This equation describes a density profile which has a weaker decrease for larger heights in comparison to Equation~(\ref{eq:isothermal_gconst}), consistent with the effective scale height increasing with $r$ due to the decreasing gravitational acceleration.

\begin{figure}[ht!]
\plotone{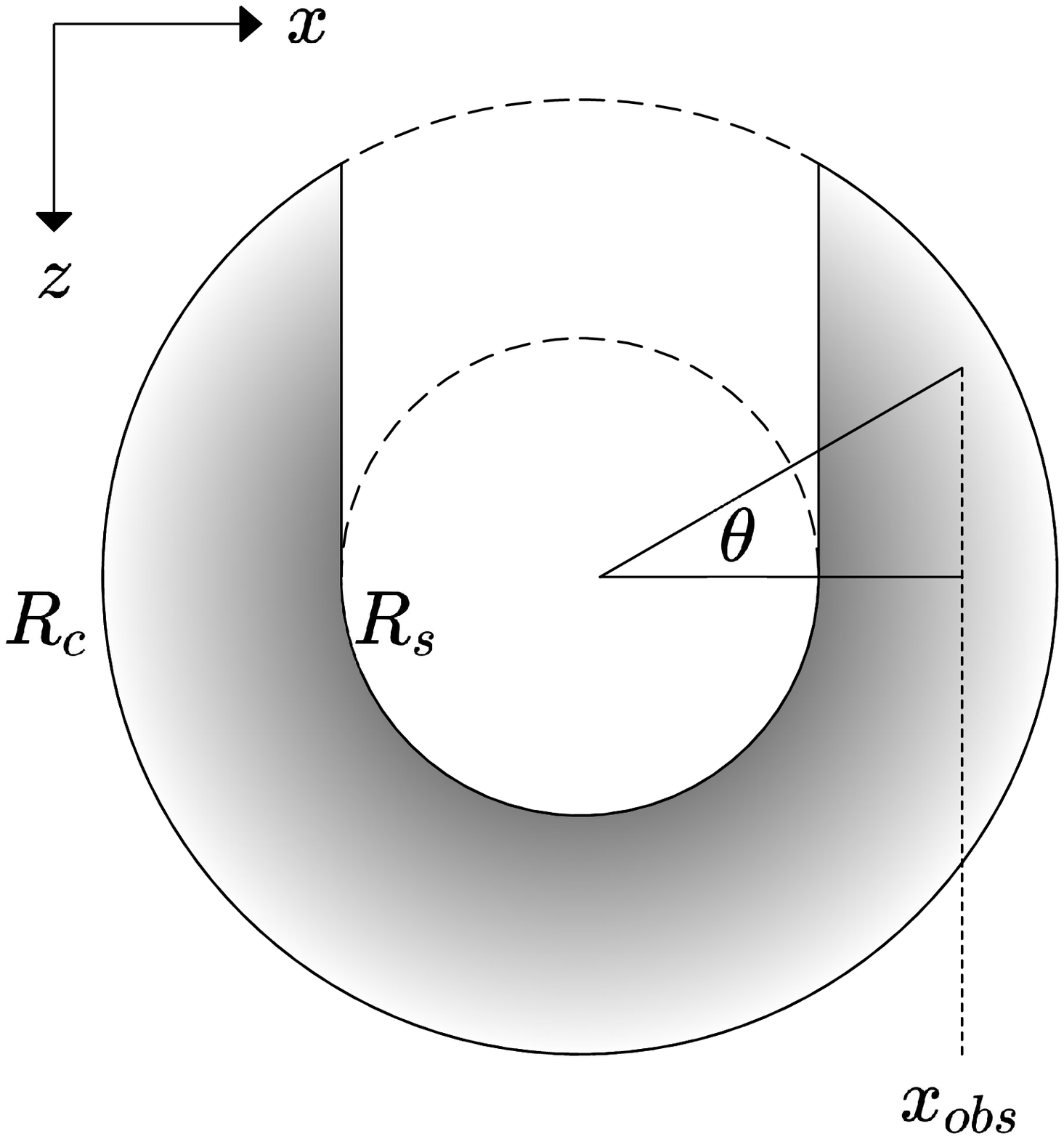}
\plotone{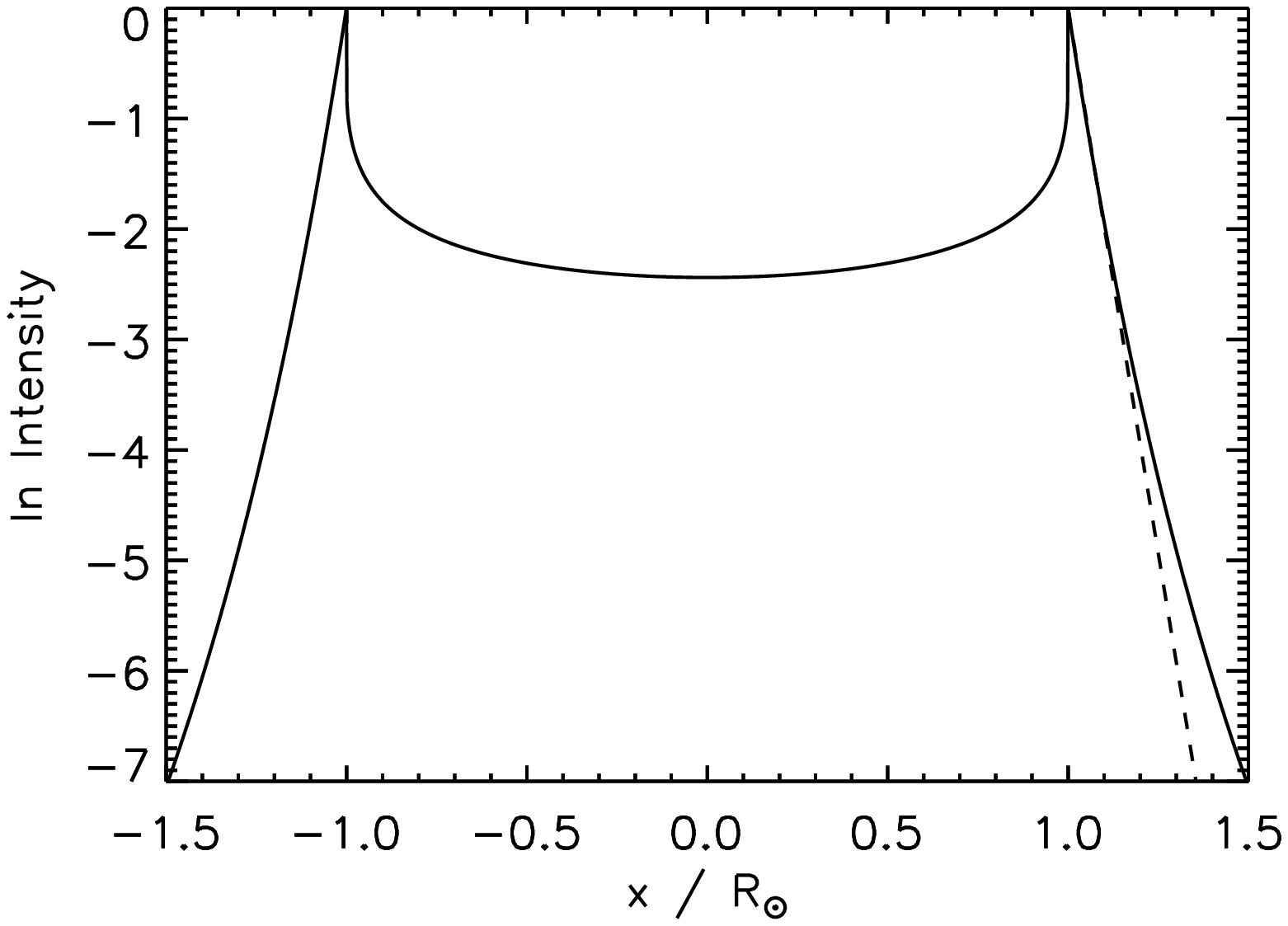}
\caption{The solar corona is modeled by an annulus between $R_{s}$ and $R_{c}$.
The observational LOS is in the $z$-direction and the intensity of optically thin EUV emission is determined by all parts of the annulus except the dashed segment obscured by the solar disk.
The bottom panel shows the logarithm of the normalised intensity profile for an isothermal atmosphere described by Equation~(\ref{eq:isothermal}).
For comparison, the dashed line represents an exponential decrease.}
\label{fig:corona}
\end{figure}

The corona is optically thin for EUV emission and so the intensity is determined by all contributions along the observational LOS.
Figure~\ref{fig:corona} shows our model for the geometry of the solar corona with regards to the EUV emission.
At a given azimuth $\phi$ (shown here for the equator $\phi=0$ so that the transverse direction is $x$), the corona is modeled by an annulus beginning at the effective solar surface $R_{s} \approx R_{\odot}$, being the radius at which the solar plasma is too cool and dense to be transparent to EUV emission and so obscures the emission from the far side of the corona (dashed segment).
For the purpose of our calculations, the maximum radius is taken to be $R_{c}=1.5 R_{\odot}$, beyond which the contribution to the intensity is assumed to be negligible.
For example, if we consider $I \propto \rho^2$ and $\rho \propto \exp \left( -r/H_{0} \right)$ then the intensity drops to 1\% of its initial value at $r \approx 2.3 H_{0}$.

A typical intensity profile for an isothermal atmosphere with temperature $T_{0}=1.2$~MK and scale height $H_{0}=60$~Mm observed by SDO/AIA 171 is shown by the bottom panel of Figure~\ref{fig:corona}.
This calculation is based on the numerical integration of EUV emission calculated for a large number of radial shells in the interval $r = \left[ R_{s}, R_{c} \right]$, corresponding to the surface and effective outer boundary of the corona, respectively.
The intensity of EUV emission at particular wavelength $\lambda$ observed at $x$ is calculated as

\begin{equation}
I \left( \lambda, x \right) = \int_{r=R_{s}}^{R_{c}} \epsilon \left( \lambda, n_e \left( r \right), T \left( r \right) \right) d_\mathrm{LOS} \left( x \right) dr
\label{eq:los}
\end{equation}
where $\epsilon$ is the monochromatic emissivity and $d_\mathrm{LOS}$ is the depth of each shell along the LOS.
Above the limb the observed intensity drops off roughly exponentially with a scale height $H_{I} \approx 0.6 H_{0}$.
(The dashed line indicates the departure from an exponential profile due to a decreasing gravitational acceleration at larger heights.)
This intensity scale height is larger than $H_{0}/2$ suggested by $I \left( r \right) \propto \exp \left( -2r/H_{0} \right)$ due to the effect of line of sight integration.
Figure~\ref{fig:corona} shows that for EUV intensities observed at a particular position $x_{obs}$ there are contributions from plasma in all radial shells satisfying $x_{obs} = r \cos \theta$, except where obscured by the solar disk.
The presence of the solar disk reduces the LOS integration depth by a factor of two for $x_{obs} < R_{s}$.
Above the solar disk the dependence is no longer well approximated by an exponential profile.

The intensity profile in Figure~\ref{fig:corona} is calculated for an effective solar surface $R_{s}=R_{\odot}$.
Since our model is concerned with EUV emission, our effective solar surface corresponds to the transition region or upper limit of the chromosphere.
The plasma in the photosphere and chromosphere is too cool and dense to produce and be optically transparent for EUV emission.
While this layer is thin in comparison with the solar radius, its presence is evident in the data.
For example, the maps of EUV emission in Figure~\ref{fig:maps} show the peak intensity at $r \approx 1.01 R_{\odot}$ (i.e. approximately $7$~Mm above the nominal solar radius).

It is desirable to include this transition region in our model.
However, our EUV data is not suitable to accurately describe this region.
We therefore approximate the presence of this opaque layer with a linear temperature profile between the photosphere and the EUV corona.
The modified temperature profile for an isothermal corona is given by

\begin{equation}
T \left( r \right) = \left\{ \begin{matrix}
\left( r - R_{\odot} \right) \frac{T_{0}-T_{\odot}}{R_{s}-R_{\odot}} + T_{\odot}, &r \le R_{s} \\
T_{0}, &r > R_{s} \\
\end{matrix}\right. ,
\label{eq:transition}
\end{equation}
where $R_{s}$ is a fitted parameter of our model along with the coronal temperature $T_{0}$, and
$T_{\odot}=5777$~K is taken to be the effective temperature of the photosphere.
We also note that the CHIANTI emissivity tables used in our forward modeling do not contain data for temperatures below $10^4$~K.

We may consider a similar layer describing the rapid decrease in density from the chromosphere to the corona. However our tests which included such a layer did not show any improvement in the quality of fit to observational data over the model without. This implies that the change in temperature occurs higher up in the atmosphere than the change in density, such that the dense plasma lower in the atmosphere is already sufficiently cool that it does not contribute to the EUV emission.

Our {\lq\lq}isothermal{\rq\rq} model for the corona is comprised of the temperature profile given by Equation~(\ref{eq:transition}) and the density profile given by Equation~(\ref{eq:isothermal}).
In Section~\ref{sect:fwdmod} we consider two additional models for the coronal density and temperature profiles.
In our {\lq\lq}exponential{\rq\rq} model we replace the constant temperature at $r > R_{s}$ with a profile of the form $T_{0} \exp \left[ k_{T} \left( r - R_{s} \right) \right]$.
In a third model, the profiles of the coronal density and temperature are permitted to vary more freely, being determined by the spline interpolation of values of density and temperature at several references heights.

\section{Forward modeling SDO/AIA EUV emission}
\label{sect:fwdmod}

For a given set of radial density and temperature profiles the emissivity of the plasma is calculated using CHIANTI tables and the response function for the particular instrument and wavelength.
This technique is adapted from the forward modeling code FoMo \citep{2016FrASS...3....4V}, with the difference being our densities and temperatures are defined by our coronal models rather than, for example, the output of numerical simulations.
The observed EUV intensity of the quiet corona is forward modeled using the plasma emissivity and the geometrical model shown in Figure~\ref{fig:corona}.
This calculation is performed by approximating the atmosphere using a large number of shells with different radii (typically $200$).
Our basic method is therefore a 2D calculation (based on the emission as a function of $x$ and $z$) which describes 1D observational data $I_\lambda \left( x \right)$.
The effect of the point spread function for each of the SDO/AIA channels is simulated by applying Gaussian smoothing with the standard deviation provided by \citet{aiapsf}.

In this section we demonstrate an application of our model using data from the six optically-thin EUV channels of SDO/AIA.
Our observational data therefore consists of six intensity profiles, and since this number of data points is typically much larger than the number of parameters in our models we have an overdetermined system.
The model parameters which most accurately describe the observational data can therefore be determined by regression analysis.
In this paper we perform Levenberg-Marquardt least-squares fits to our data using the \textsc{MPFIT} code \citep{2009ASPC..411..251M} in \textsc{IDL}.

\subsection{Quiet corona}
\label{sect:quiet}

\begin{figure*}[ht!]
\plottwo{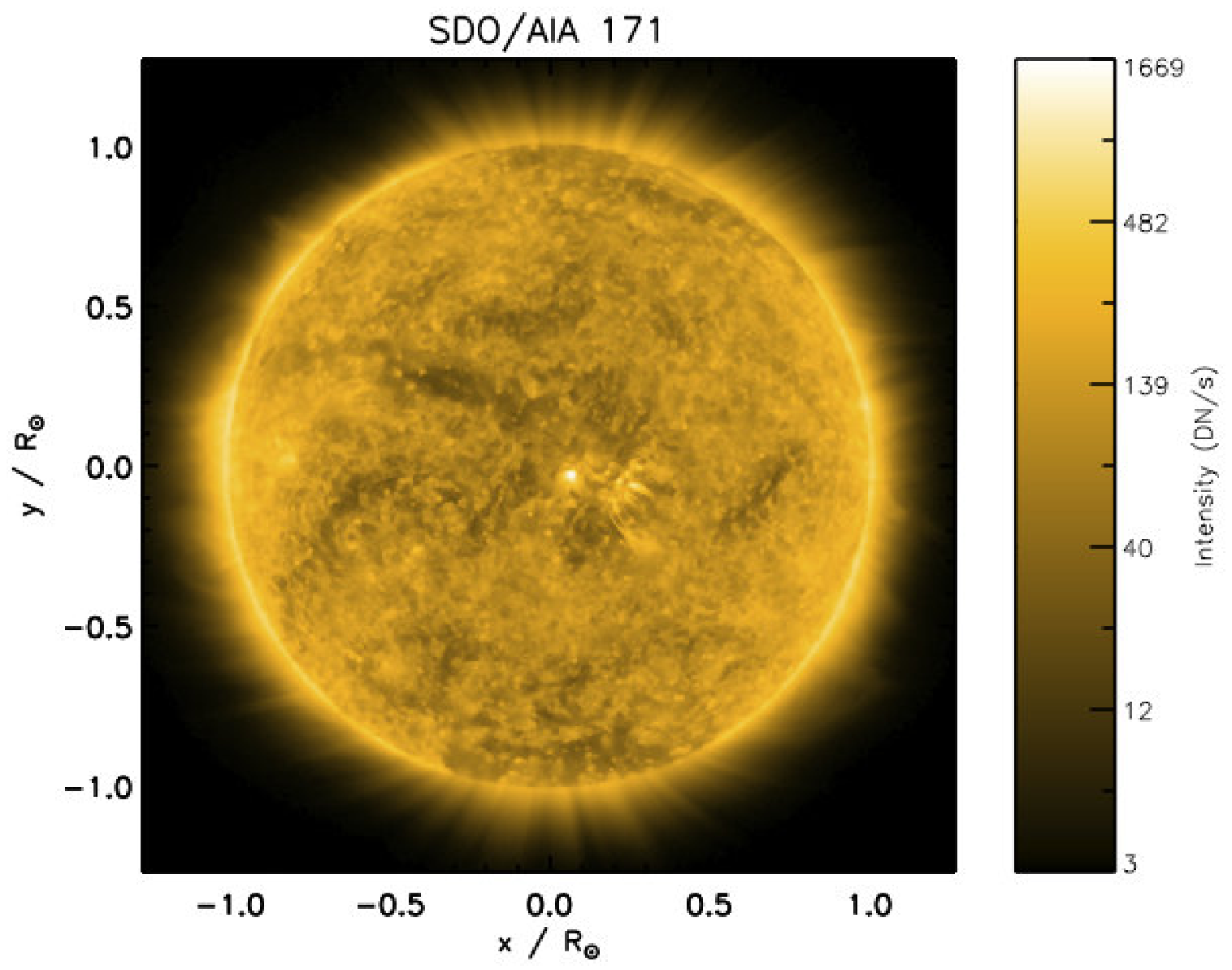}{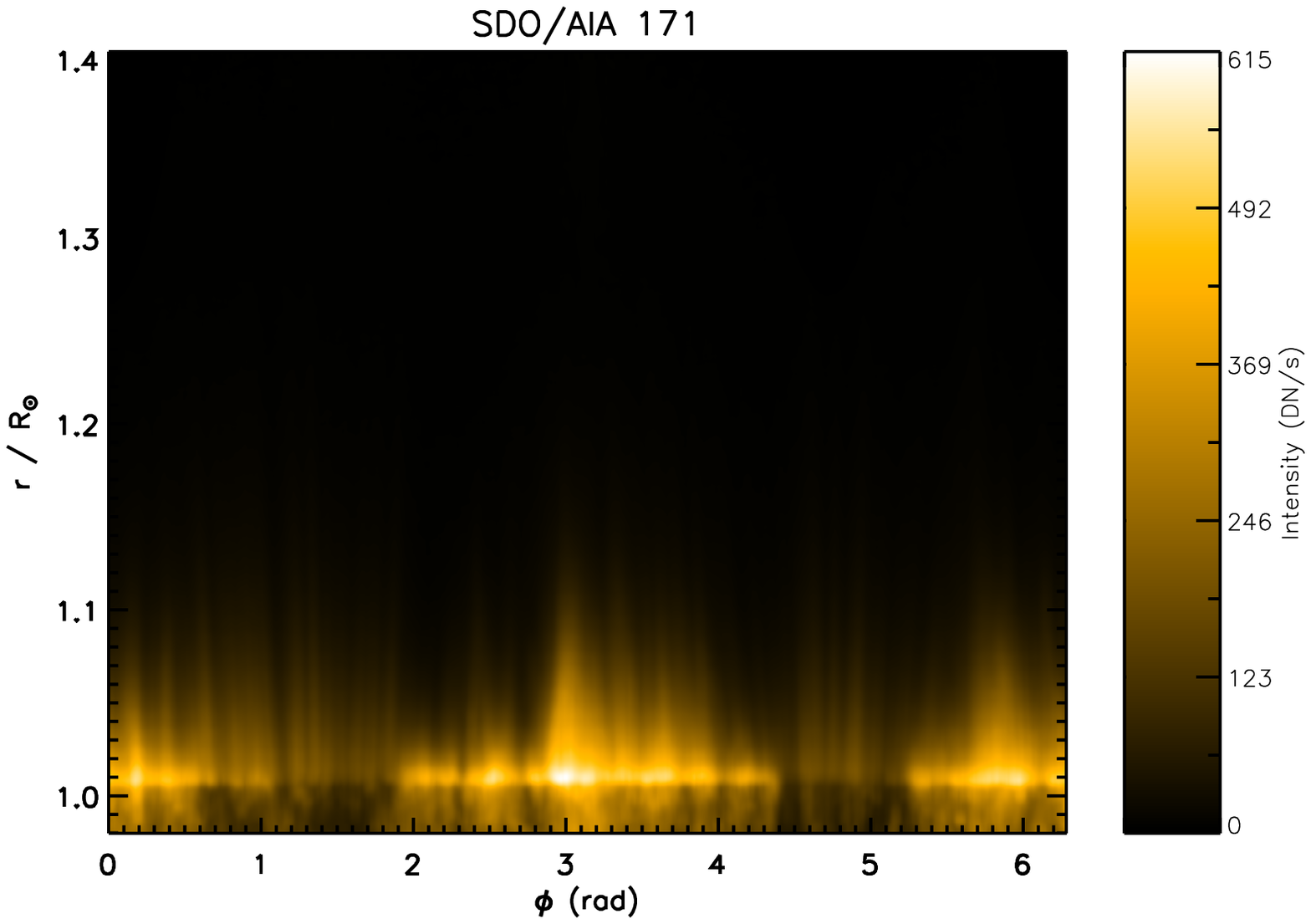}
\plottwo{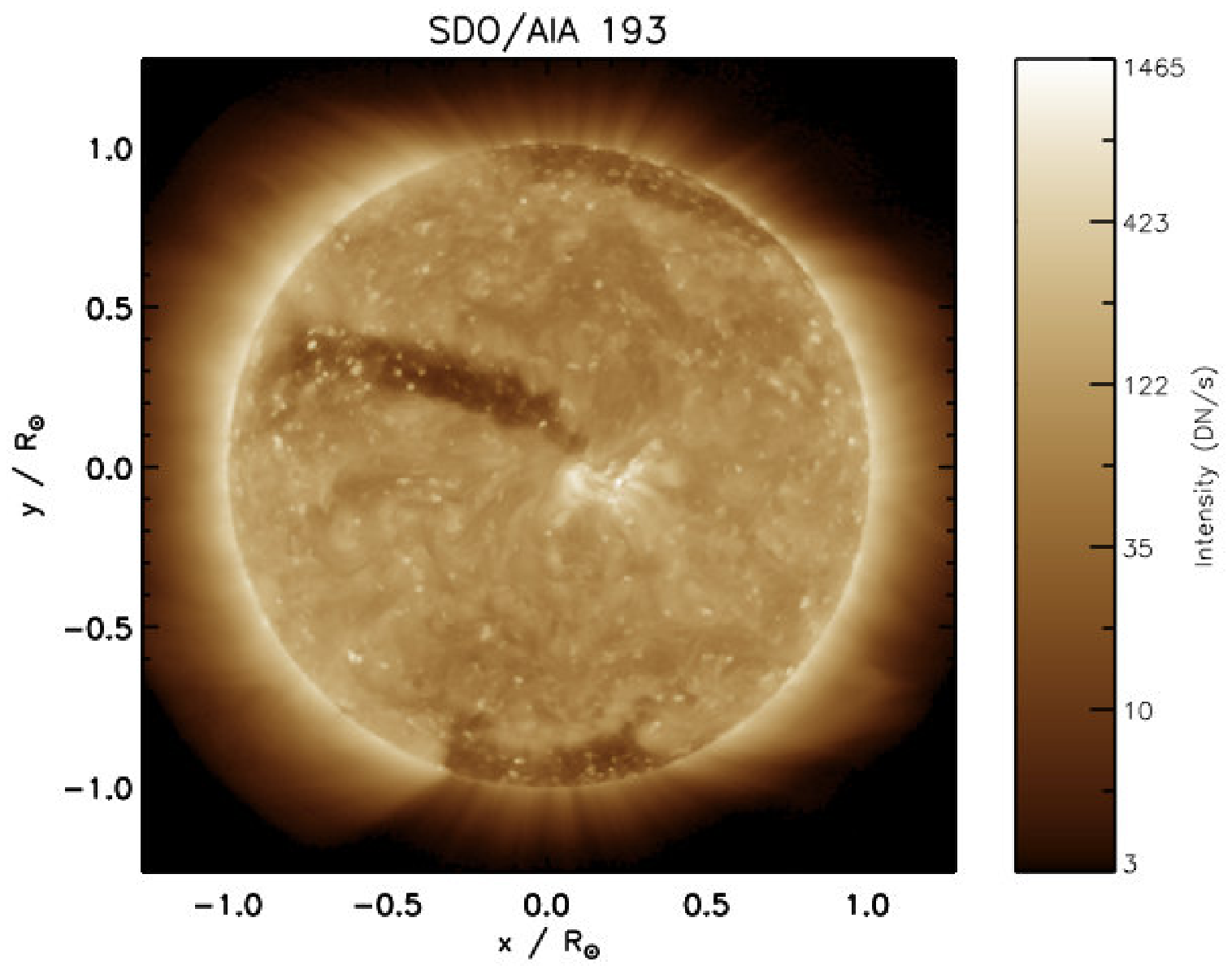}{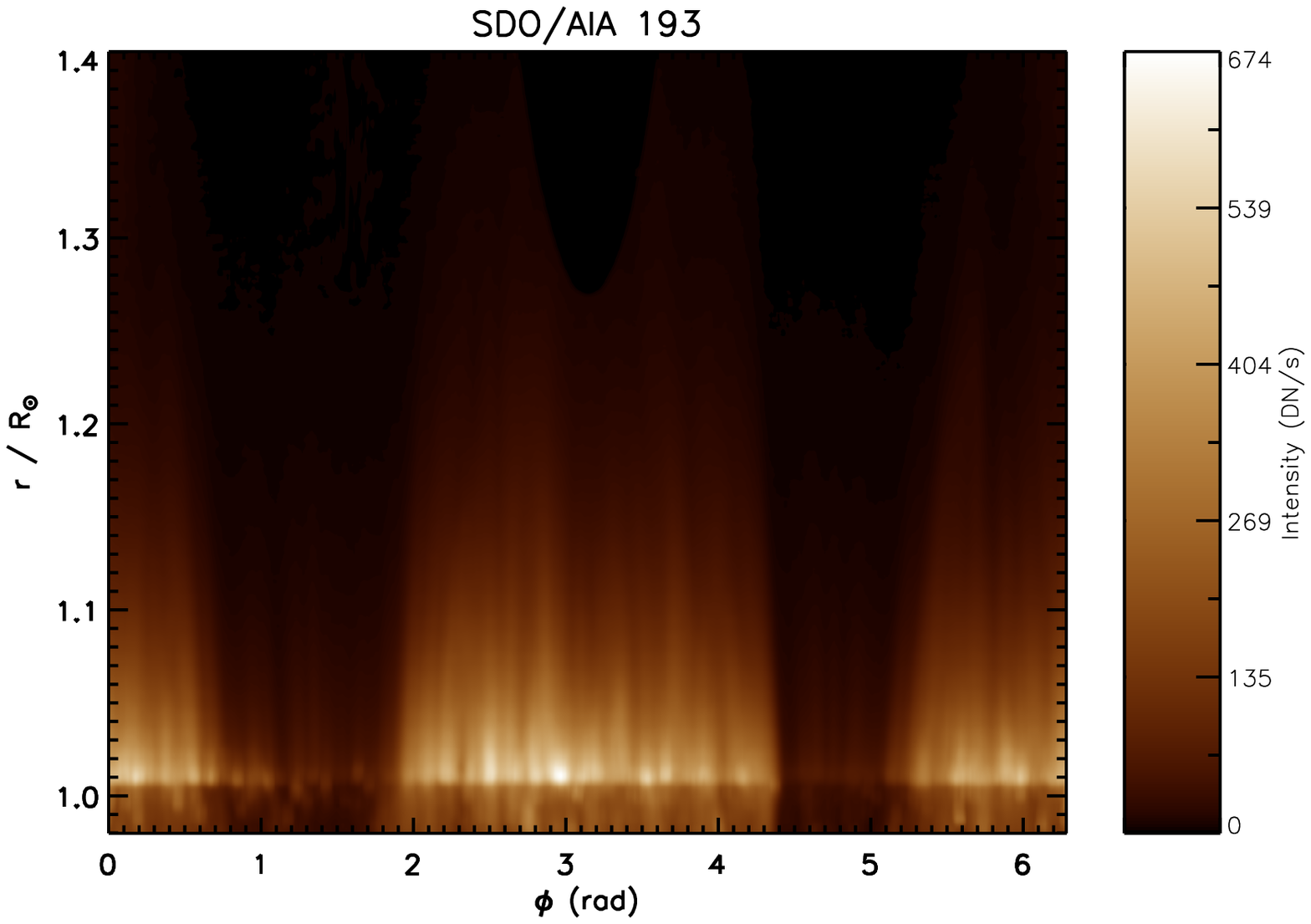}
\caption{SDO/AIA data for $171$ and $193$~{\AA} channels at 15:47:44 on 10 March 2018. The \textit{left} panels show intensity images. The \textit{right} panels show the corresponding maps of the corona, near and above the limb, in polar coordinates.}
\label{fig:maps}
\end{figure*}

We first demonstrate our method by applying our models to data taken at a time when the Sun appears relatively quiet.
Our data are taken at 15:47:44 on 10 March 2018 ($\pm 4$~seconds depending on the particular channel).
This time was chosen to minimize the number of active regions or other magnetic structures near the limb of the Sun (several are, however, present on the disk).
We use the intensity profiles above the limb to compare with our model since these regions exhibit far less variation in intensity than the solar disk.

We use the (up to) six optically thin EUV channels of SDO/AIA
($094$, $131$, $171$, $193$, $211$, and $335$~{\AA}) in our analysis, i.e. excluding the optically thick $304$~{\AA} channel.
We obtain the data as Level 1 AIA files and use the \textsc{aia\_prep} routine for additional calibration, including exposure normalization.

Since our models describe the radial profiles of density and temperature it is convenient to process our data using polar coordinates
$\left( x', y' \right) = \left( r' \cos \phi', r' \sin \phi' \right)$,
where the prime symbol denote observational (projected) coordinates rather than physical ones.
The images and coronal maps for our observation are shown in Figure~\ref{fig:maps} for two of our six channels.
The presence of super-radially expanding polar coronal holes is evident, particularly in the $193$~{\AA} channel.

Our forward modeling for each of the six channels is based on the instrument response function calculated for when SDO began observations in 2010. The sensitivity of each AIA channel has degraded over time \citep{2012SoPh..275...41B}.
We account for this degradation by applying the appropriate correction factors to each of our intensity profiles based on the peak response for each channel at the time of the observation, which is obtained using the \textsc{aia\_get\_response} procedure.

\begin{figure*}[ht!]
\plottwo{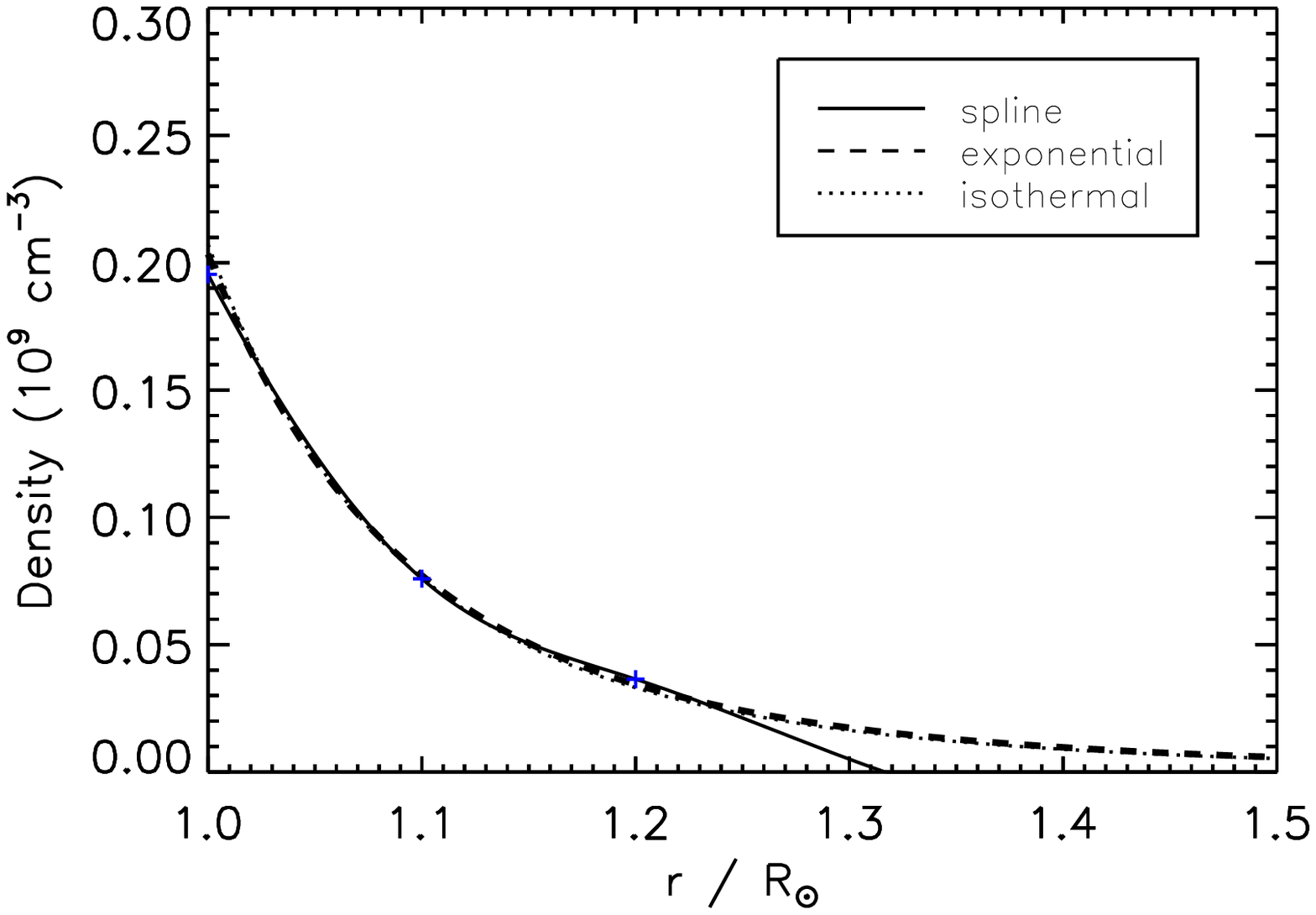}{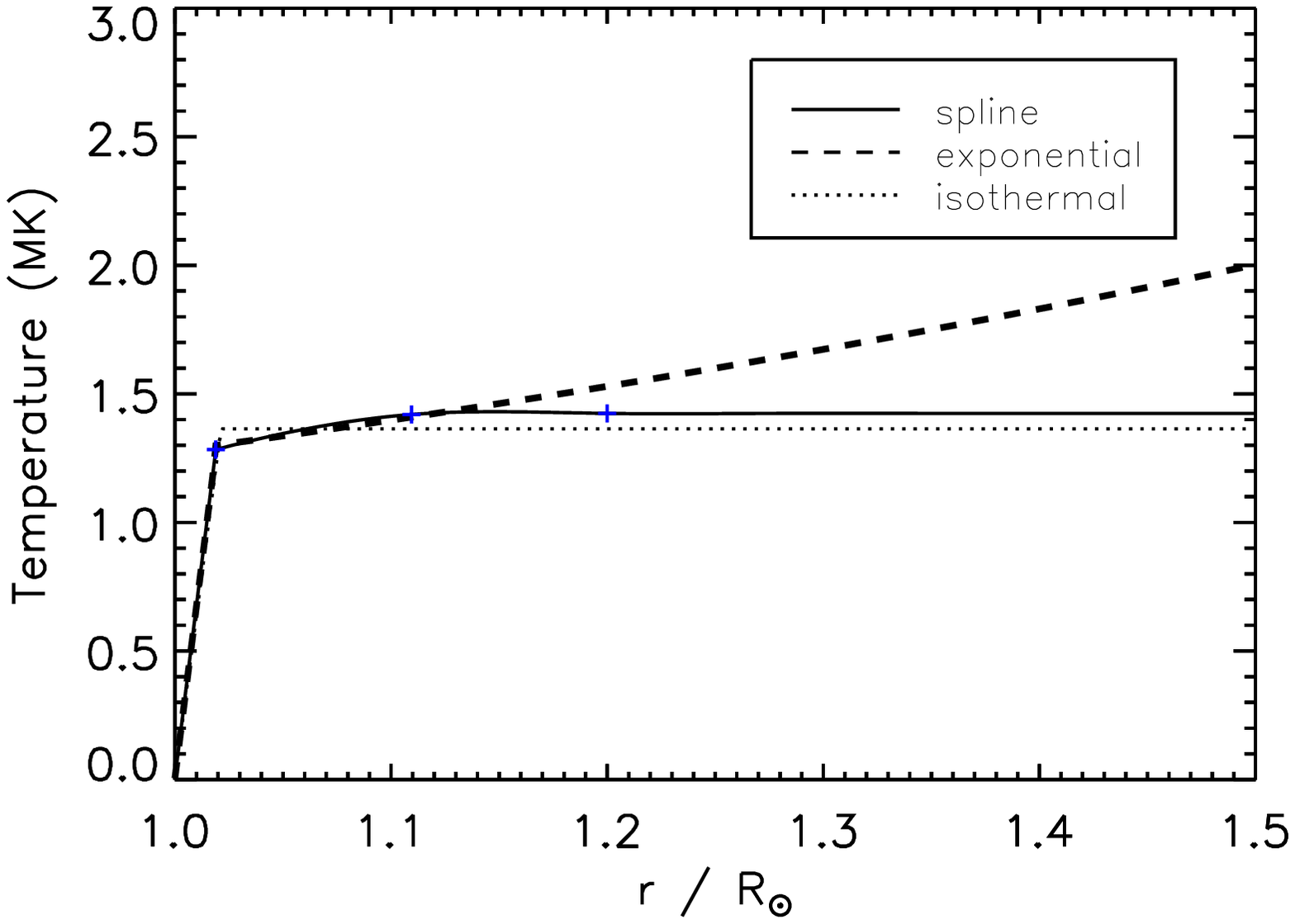}
\caption{Density and temperature profiles for our model fits.
The solid, dashed, and dotted lines correspond to our spline, exponential, and isothermal models, respectively.
The blue crosses denote the fitted spline interpolation points.}
\label{fig:model_fits}
\end{figure*}

\begin{figure*}[ht!]
\plottwo{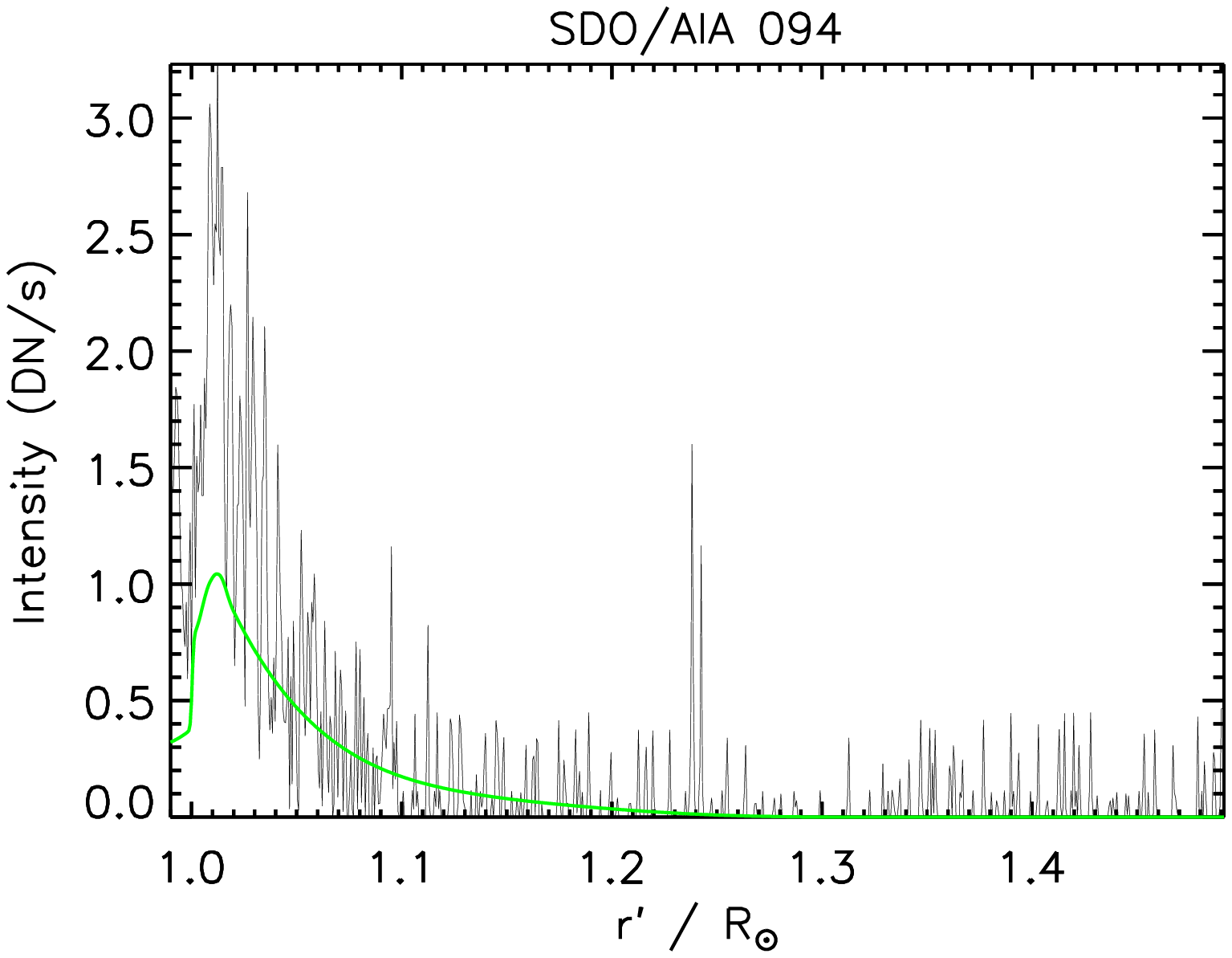}{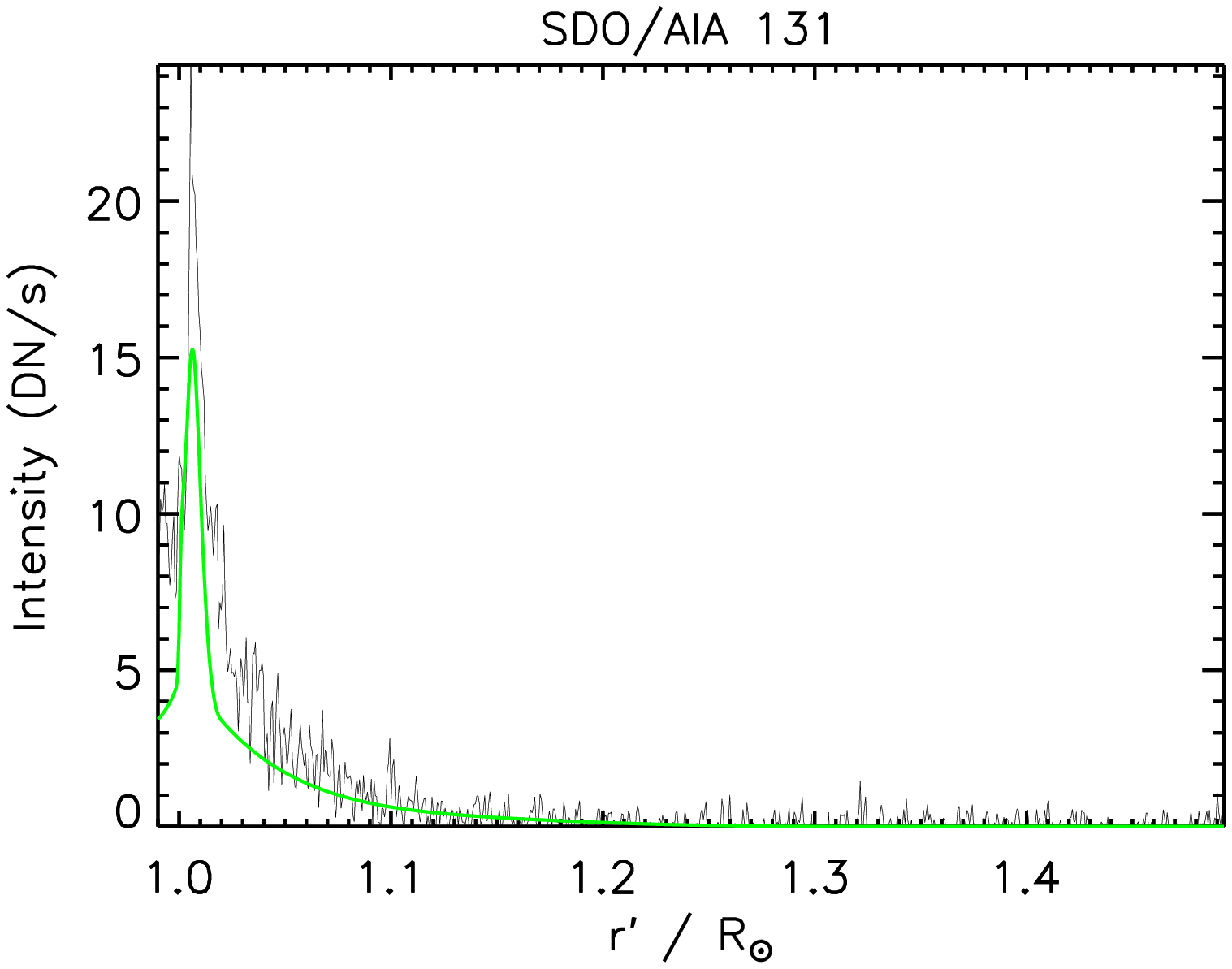}
\plottwo{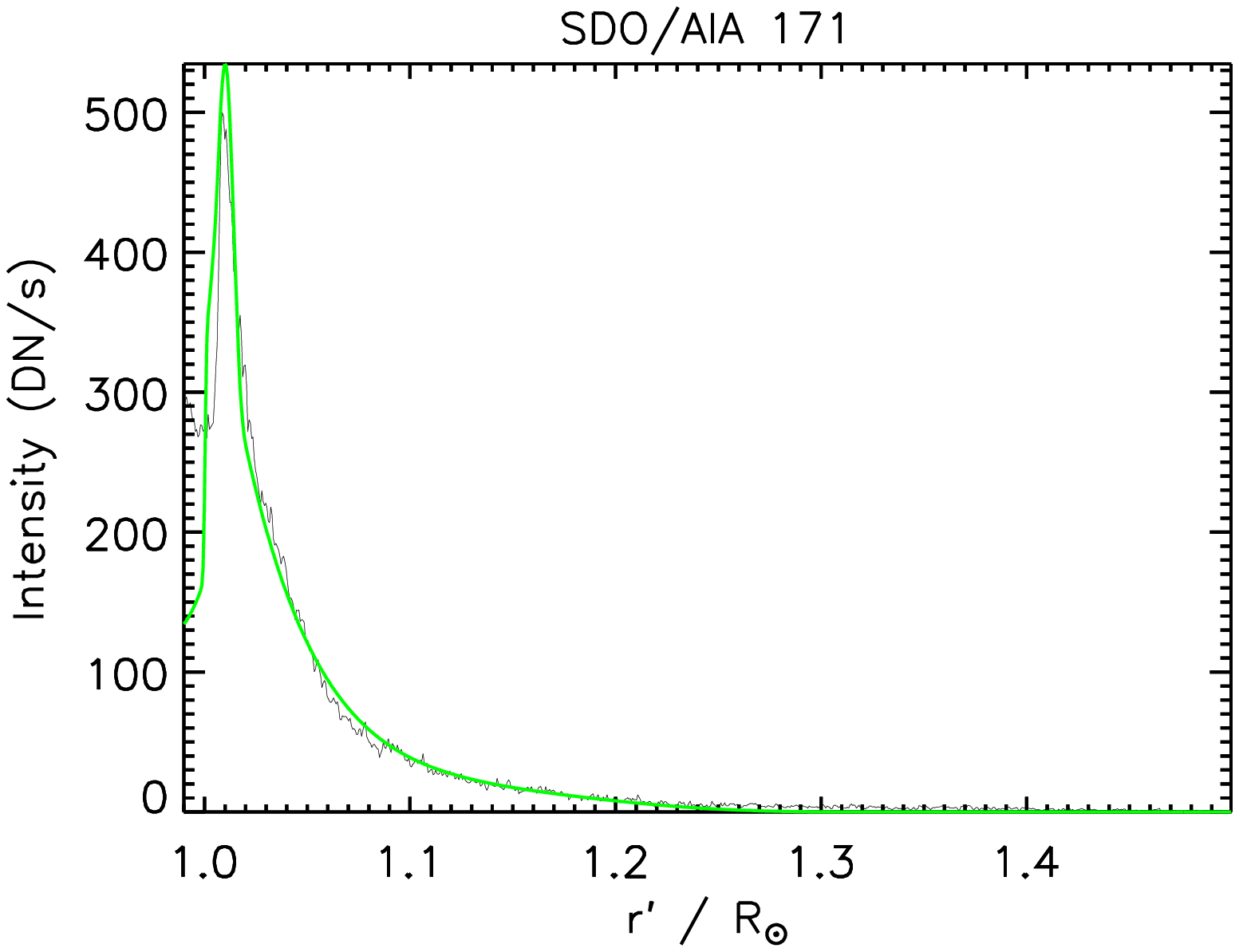}{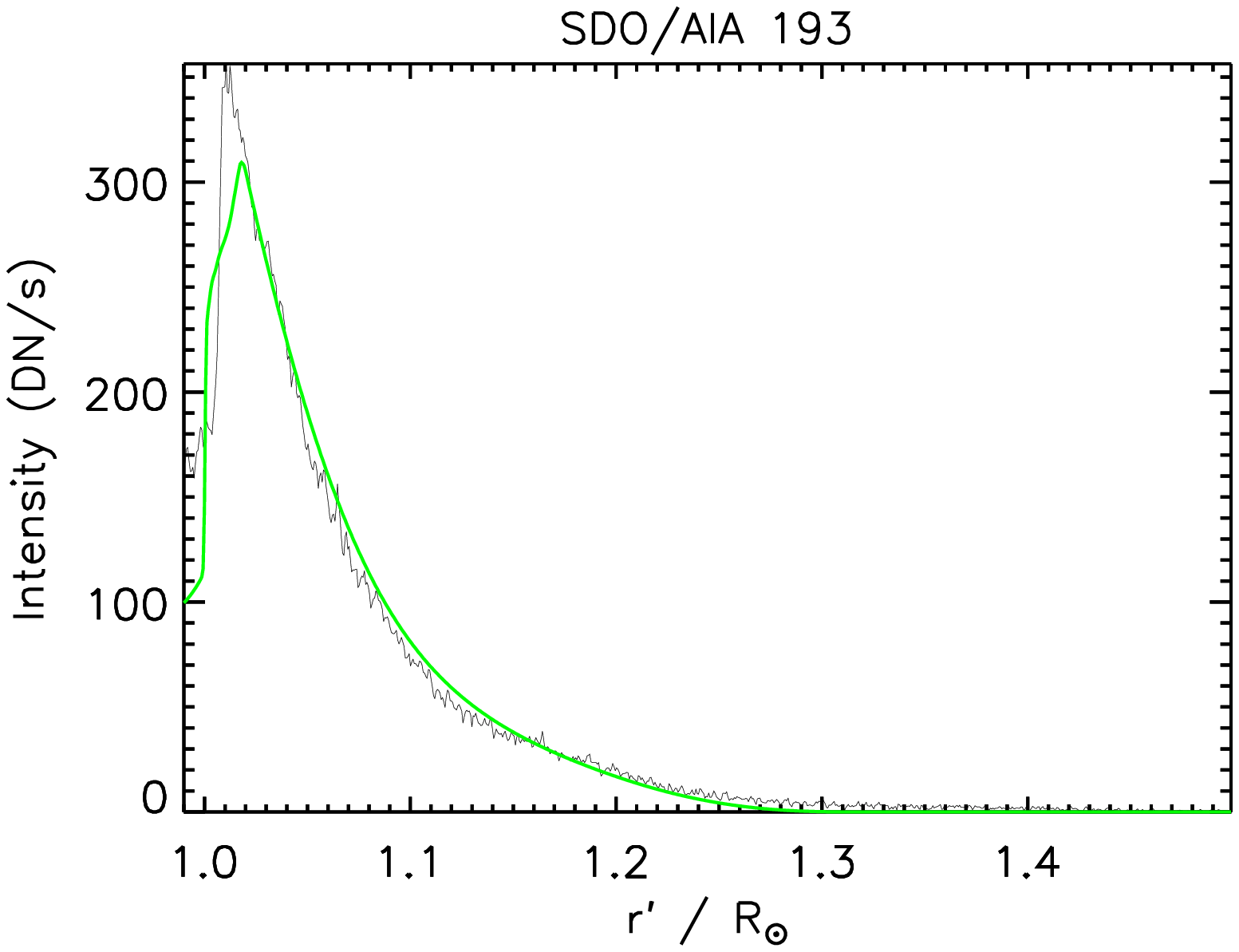}
\plottwo{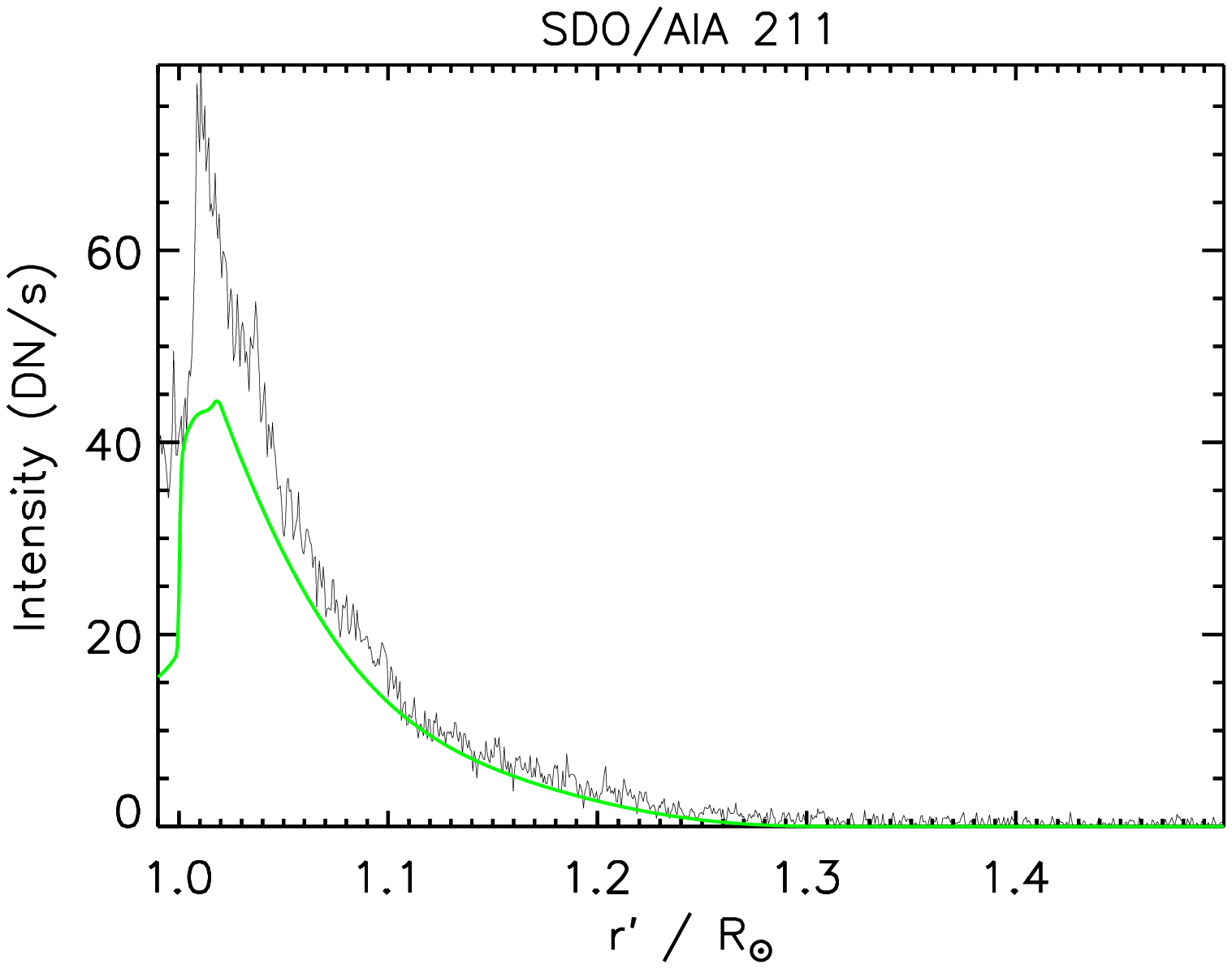}{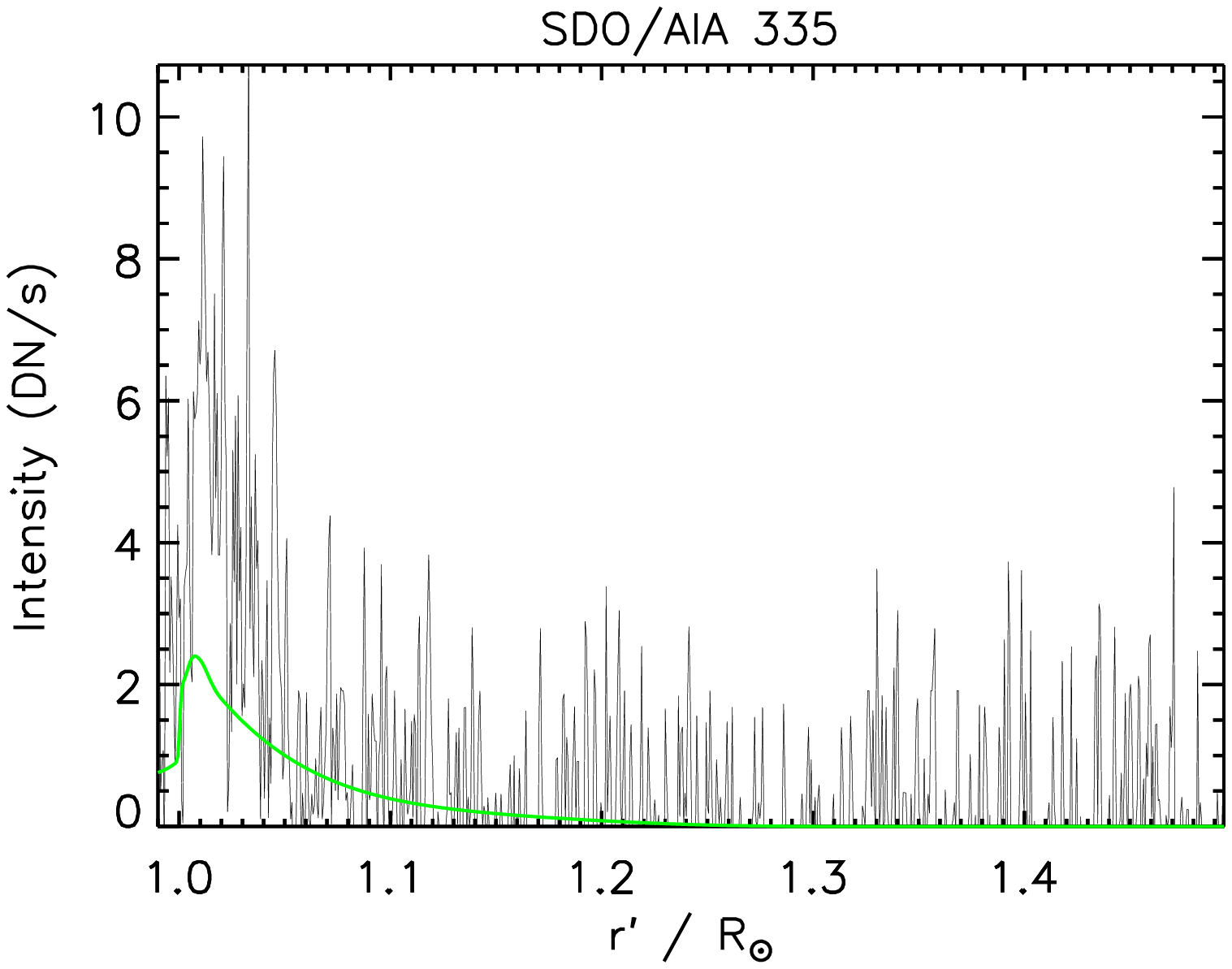}
\caption{Results of our spline-based model for $\phi'=5.5$~rad for the intensity profiles for each of the six EUV AIA channels.
The green lines show the model intensities.}
\label{fig:spline_channels}
\end{figure*}

\begin{figure}[ht!]
\plotone{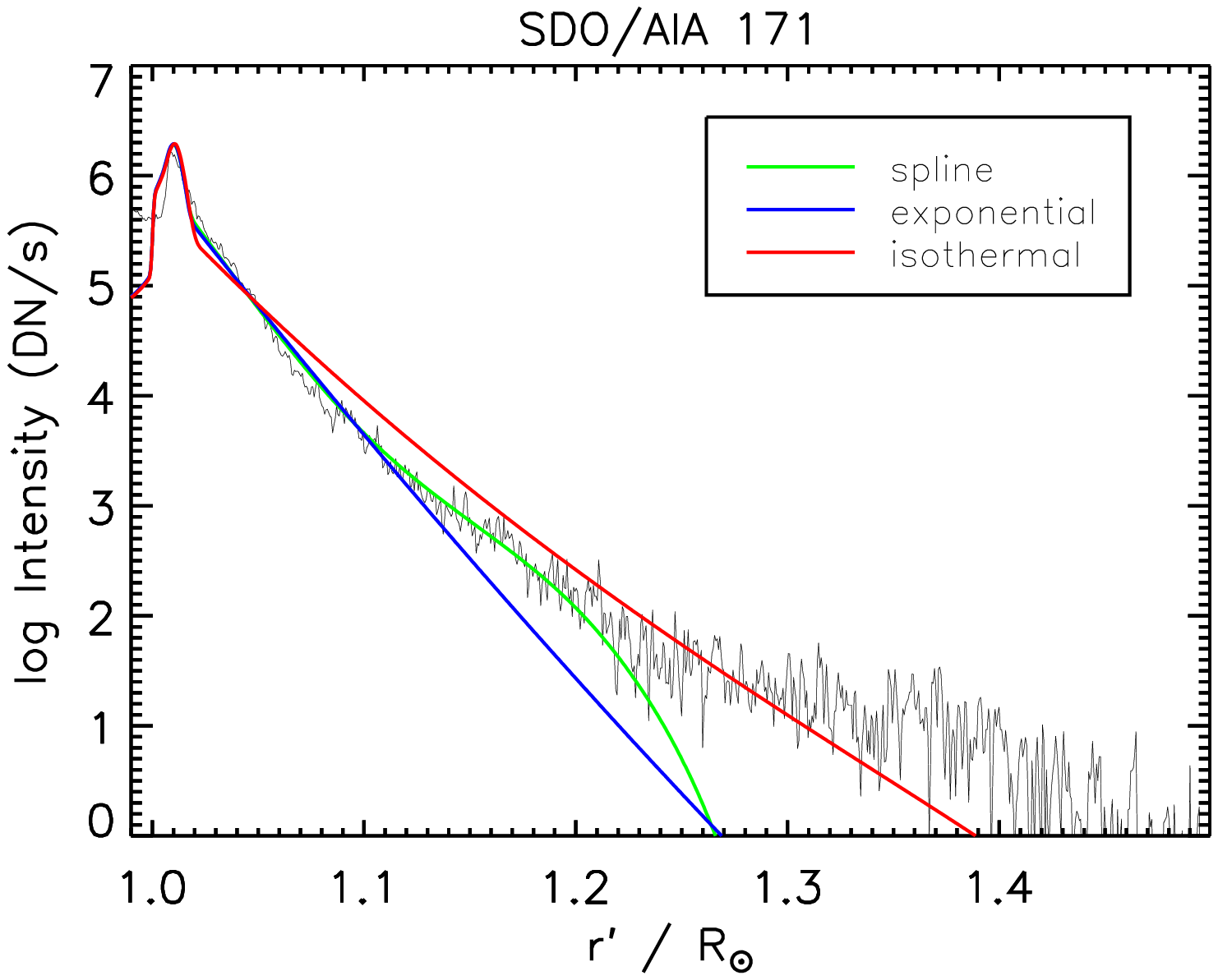}
\plotone{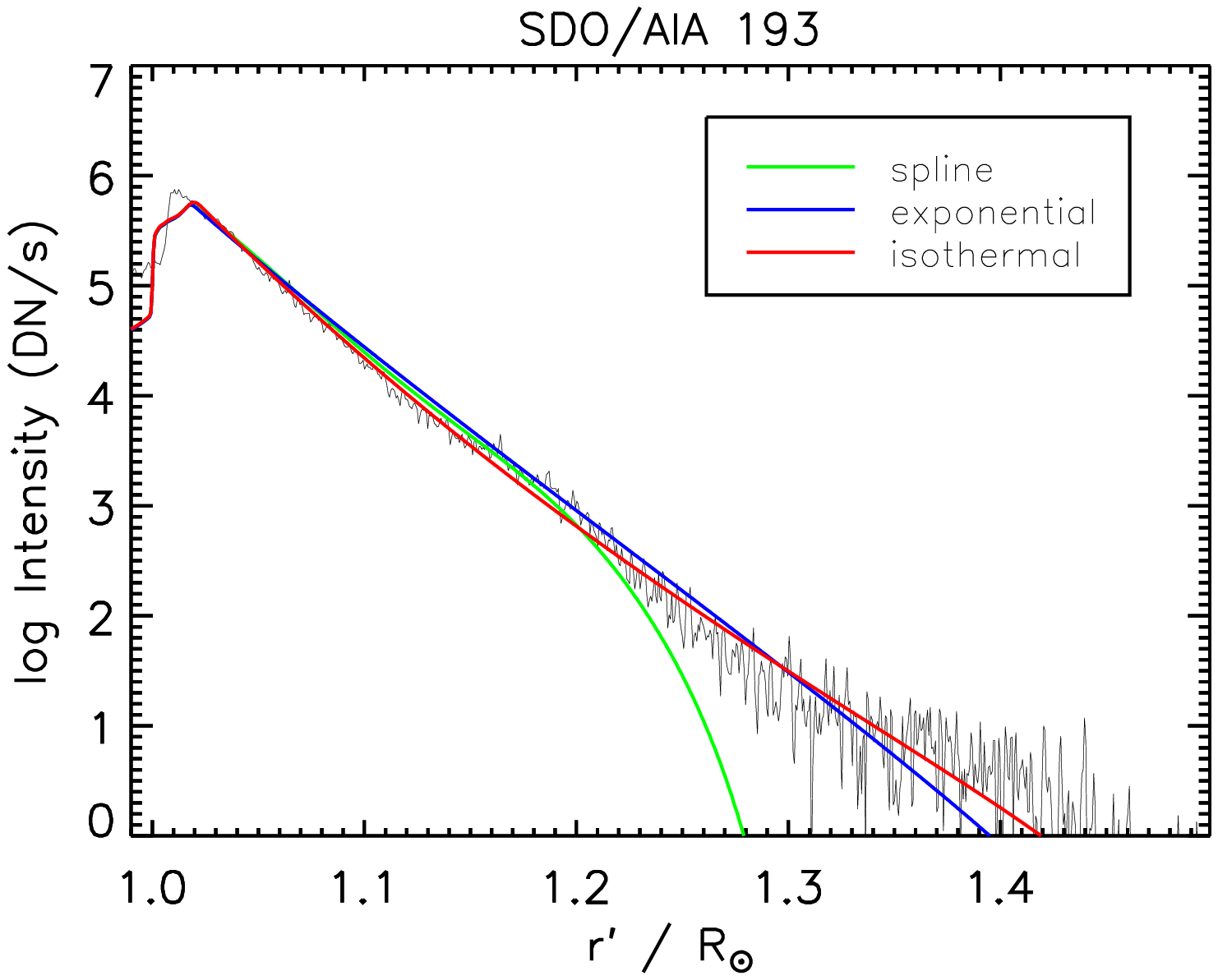}
\plotone{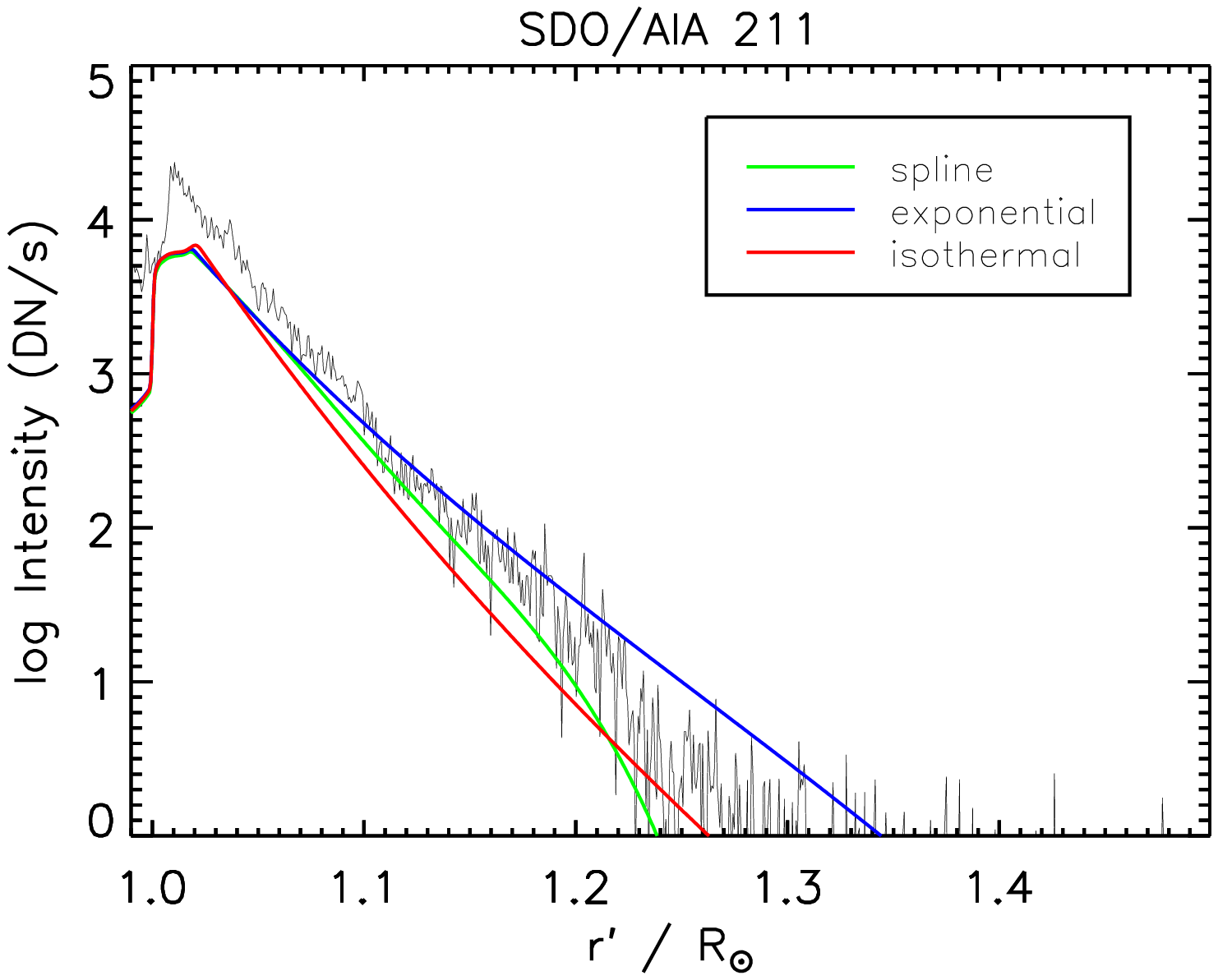}
\caption{Comparison of the forward modeled intensity for our three models for the $171$, $193$, and $211$~{\AA} channels.}
\label{fig:compare_models}
\end{figure}

Figure~\ref{fig:model_fits} shows the density and temperature profiles calculated for $\phi'=5.5$~rad.
The dotted lines correspond to our isothermal model given by Equations~(\ref{eq:isothermal}) and (\ref{eq:transition}).
The dashed lines are for our exponential model which is the same except for allowing the temperature to vary with an exponential profile above $R_{s}$.
The best fit suggests a temperature which is increasing with height.
Since the intensity of EUV emission decreases with height we must consider that the results will be most strongly influenced by the behaviour near to the limb.
The solid lines correspond to our model based on spline interpolation of values at several heights. Here, four values are used and are indicated by the blue plus symbols.
The lowest of these interpolation points is at $R_{s}$ and the highest is at $1.2 R_{\odot}$. Above this height the density profile is determined by an additional point (zero density at $1.5 R_{\odot}$) while the temperature profile is assumed to remain constant.
Our spline model produces equally good or better results than the exponential model and so the increasing temperature with height inferred by our exponential model should only be considered to apply to $r \lesssim 1.2 R_{\odot}$.

Figure~\ref{fig:spline_channels} shows the data and model fits for each of the AIA channels using our spline model.
The intensities are negligible with high noise in the very high temperature $094$ and $335$~{\AA} channels since we consider the background corona rather than flaring emission.
We see that our model is able to simultaneously reproduce the six EUV intensity profiles reasonably well.
Figure~\ref{fig:compare_models} shows a comparison of the forward modeled intensity (with a logarithmic scale) for our three density and temperature profile models.
The reduced chi-squared values for our spline, exponential, and isothermal models are $136$, $141$, and $151$, respectively.
Our least-squared fits are performed using unweighted data. Poisson weighting would be appropriate to describe the shot noise associated with the CCD images. However we expect the systematic errors arising due to our simplified model to be more significant, in particular bright magnetic structures whose influence would be magnified by Poisson weighting.
Our spline model better reproduces the observational data owing to its greater number of fitted parameters, except for heights above $1.2 R_{\odot}$ where the EUV intensity is negligible and insufficient to constrain the model. (At these large heights the exponential and isothermal models do not require data to constrain them since their monotonic functions are simply extrapolated from the shape fitted at lower heights.)
The largest discrepancies occur near the solar limb where our assumption of optical transparency breaks down due to the presence of the chromosphere and transition region.
We can therefore consider our method to be most appropriate for distances of approximately $30$--$300$~Mm above the solar surface (using SDO/AIA data).
Calculating the gas pressure for the spline model fit produces results consistent with the hydrostatic equilibrium in Equation~(\ref{eq:hydrostatic}).

The behaviour we find is in good agreement with that calculated by \citet{2018ApJ...852..137R} using the \textsc{CODET} model (see their Figure 4).
Their method calculates the EUV emission for the $193$ and $211$~{\AA} channels but averages the intensities over the solar disk and uses scaling laws to relate the plasma density and temperature to the extrapolated magnetic field strength.
On the other hand, our approach is based only on forward modeling the spatial profiles of EUV emission with no additional physical modeling.

\begin{figure*}[ht!]
\plottwo{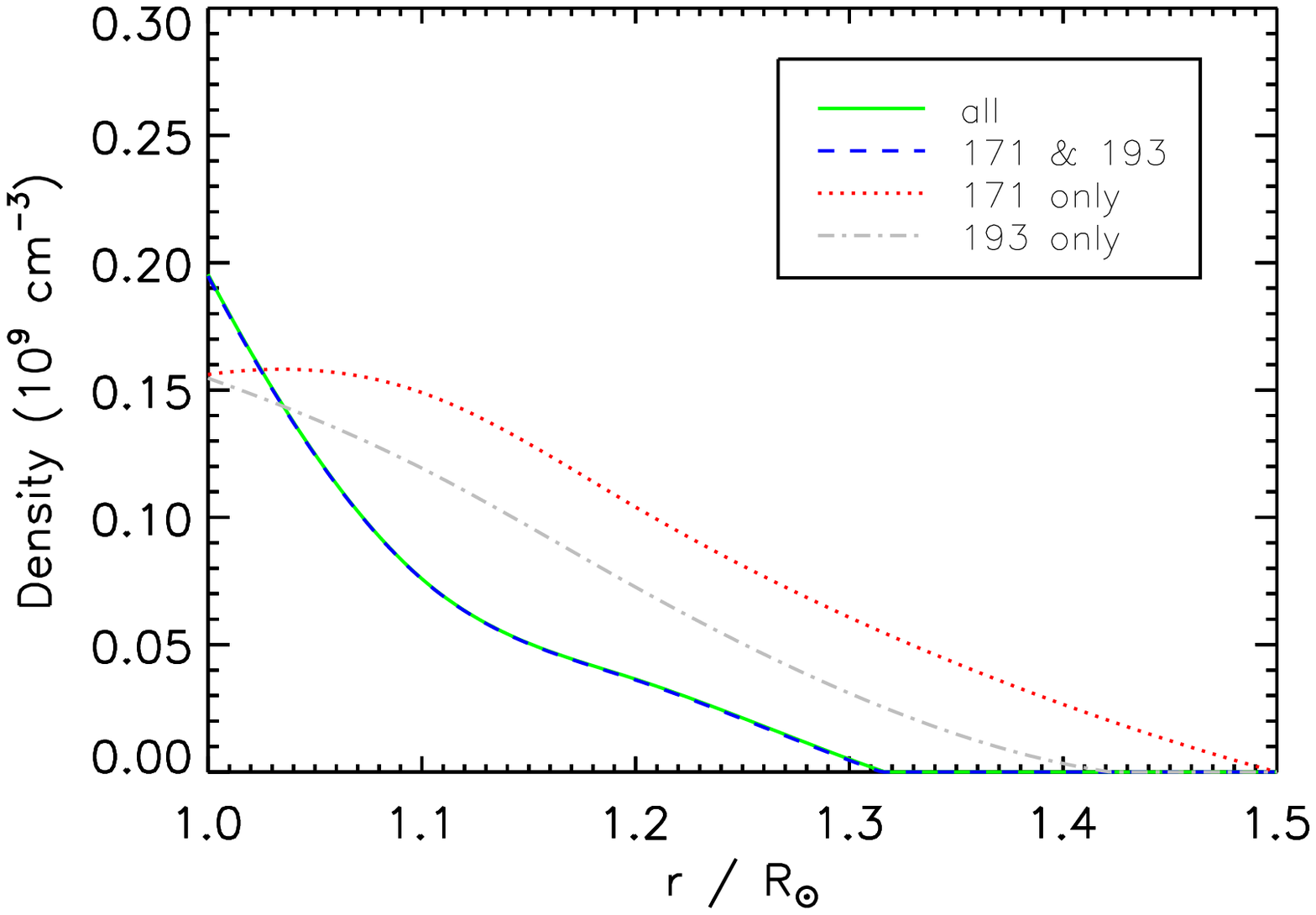}{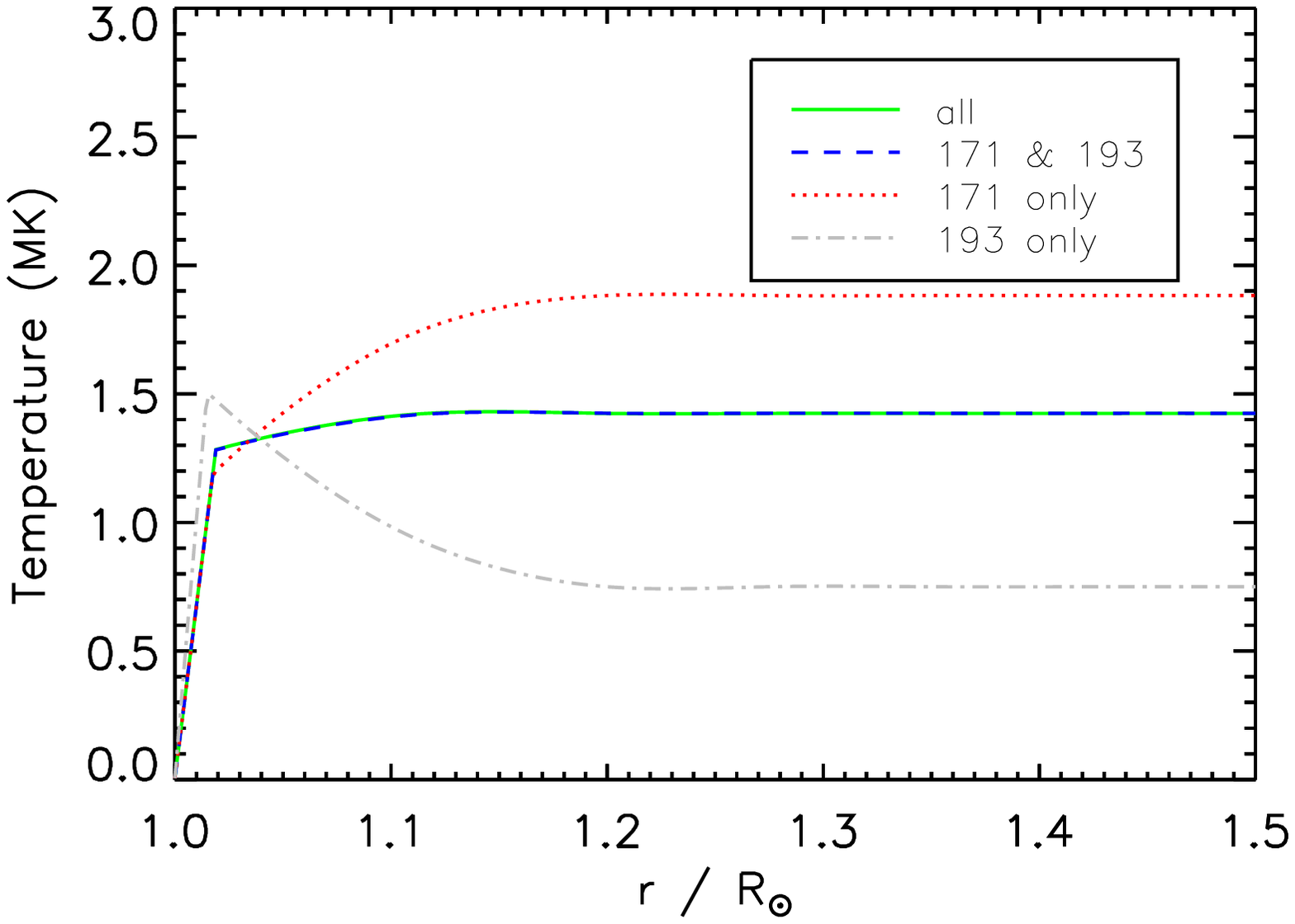}
\caption{Density and temperature profiles for our spline model using data from different numbers of the six optically thin EUV channels of SDO/AIA.}
\label{fig:channels_comp}
\end{figure*}

Since we use spatial information to constrain our models, our fitting problem is overdetermined (in contrast to differential emission measure inversions) and it is not necessary for us to use all six SDO/AIA channels. Figure~\ref{fig:channels_comp} shows a comparison on density and temperature profiles obtained by our spline model when using data from different channels. For the quiet corona, our results using the two channels with highest intensity ($171$ and $193$) are identical to when all six channels are used. However, using either of these channels on its own produces significantly different results. A minimum of two channels is required to distinguish between changes in EUV intensity arising due to changes in density and/or temperature.

\begin{figure}[ht!]
\plotone{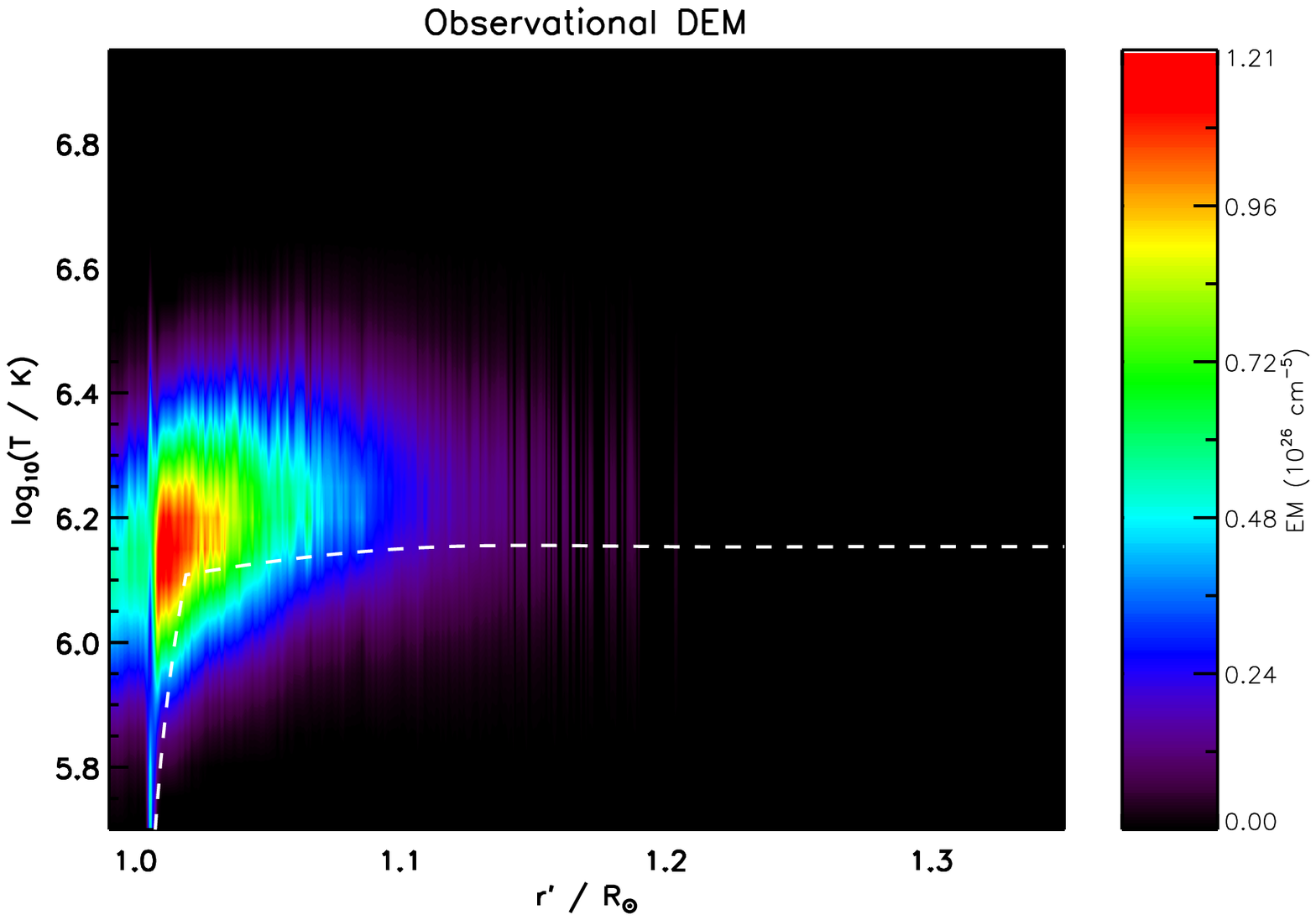}
\plotone{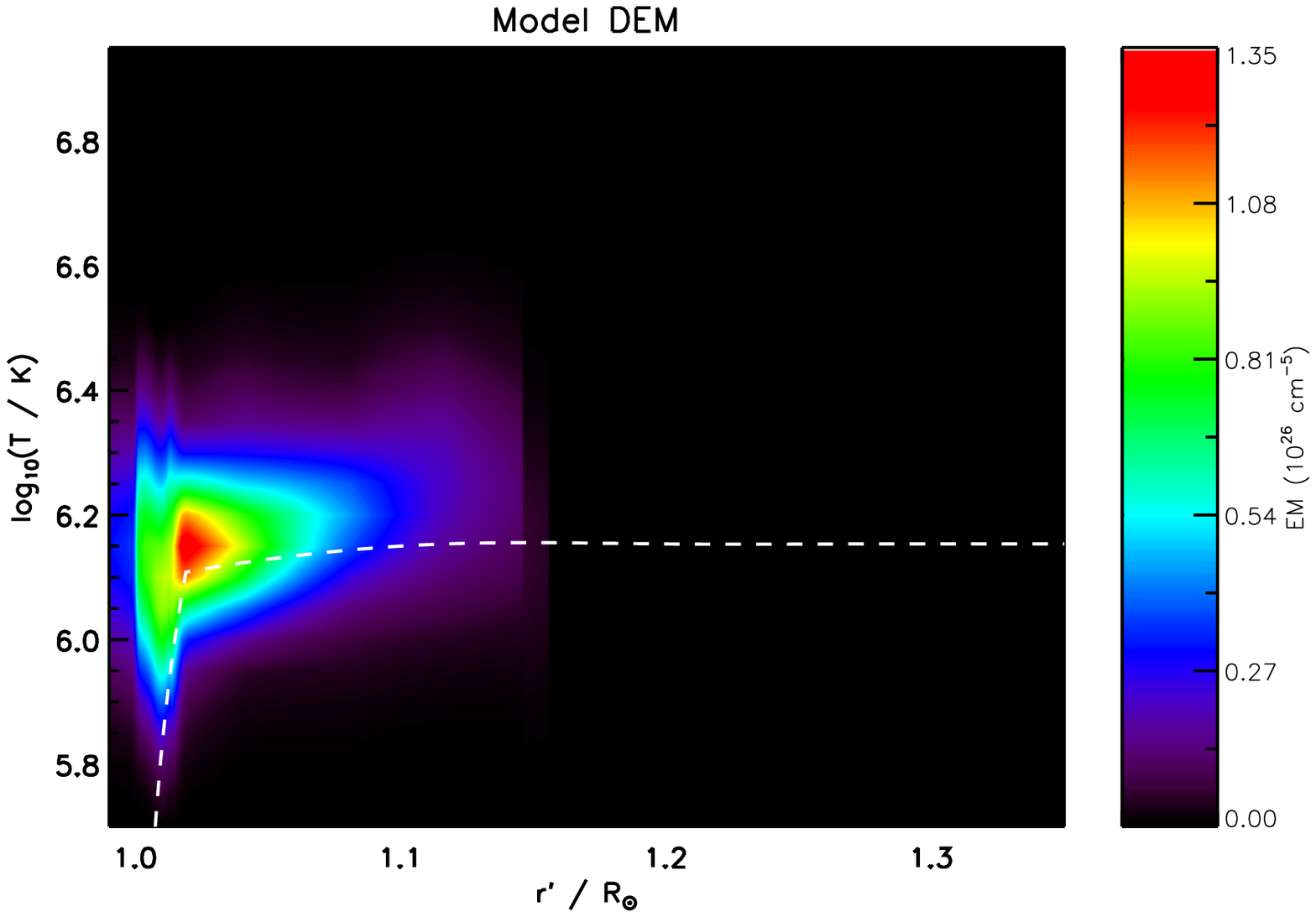}
\plotone{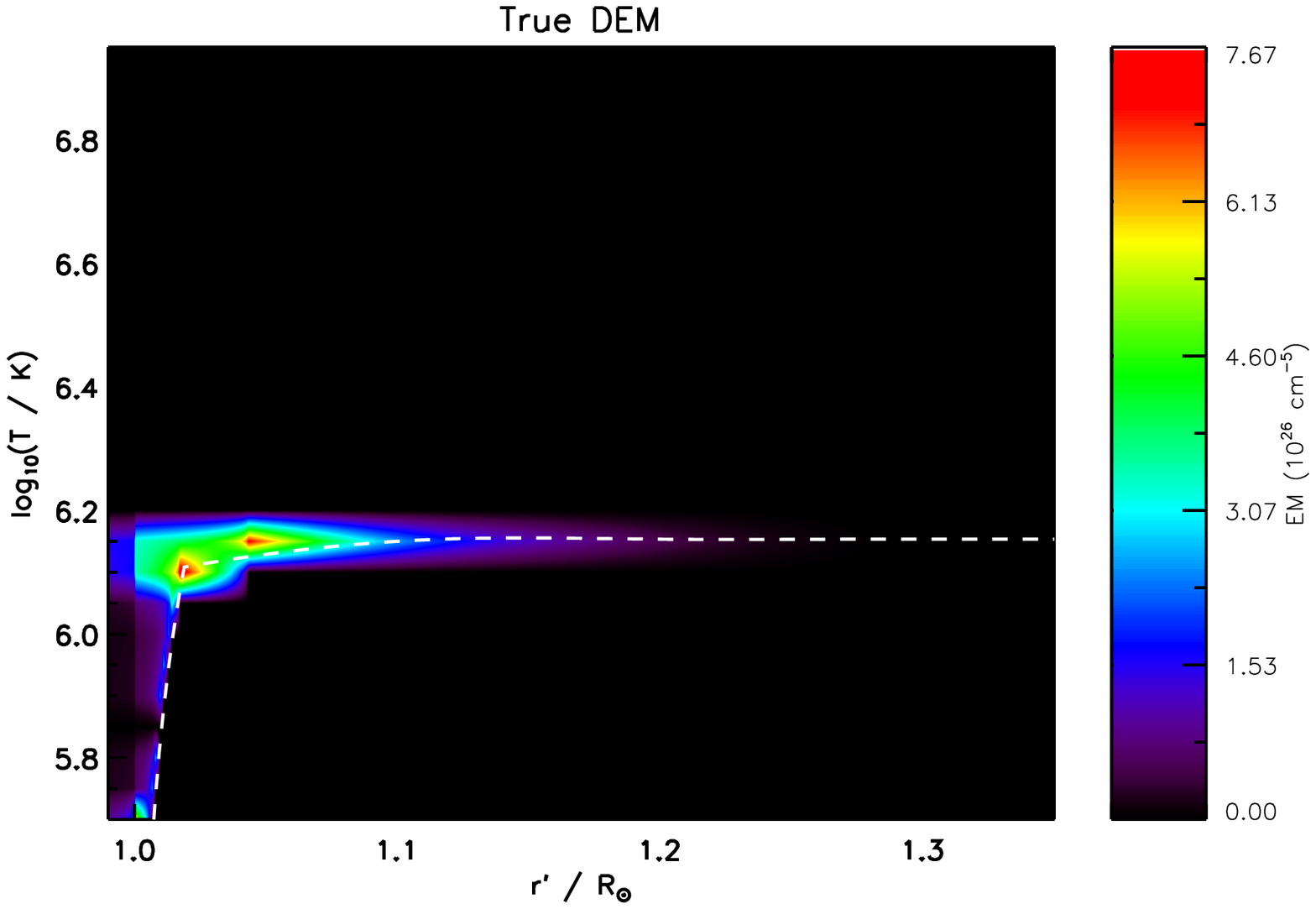}
\caption{DEM analysis for $\phi'=5.5$~rad using the sparse inversion method of \citet{2015ApJ...807..143C}.
The \textit{top} and \textit{middle} panels correspond to results for the observational and modeled intensities, respectively.
The \textit{bottom} panel shows the true DEM calculated from the model.
The white dotted lines represent the fitted temperature profile for our spline-based model.}
\label{fig:dem}
\end{figure}

It is instructive to compare our results with differential emission measure inversions since this is a common method to obtain information about the temperature of coronal plasma.
The top panel of Figure~\ref{fig:dem} shows the result of DEM analysis of our observational data the sparse inversion method of \citet{2015ApJ...807..143C}.
The middle panel shows the same method applied to our fitted model intensity profiles shown in Figure~\ref{fig:spline_channels}.
This inversion method includes corrections to account for the AIA response degradation and so we apply it to our uncorrected observational data.
(In the case of our model intensity profiles, we degrade our results using the reverse of the corrections we already applied in our forward modeling.)
The bottom panel shows the true DEM present in the model as calculated directly from the density and temperature profiles rather than inversion of the intensity profiles.
The dashed white lines is our fitted temperature profile from Figure~\ref{fig:model_fits}.
The DEM spectra are generally in good agreement, particularly when describing the range of temperatures detected just above $R_{\odot}$ where the LOS integration has contributions from the largest number of shells.
However, the model DEM obtained from inversion is significantly more broad than that from direct calculation.
This broadening effect has recently been studied by \citet{2018A&A...620A..65V} in which the authors advocate caution for interpreting broad DEMs as evidence of multi-thermal plasma.

\subsection{Coronal holes}
\label{sect:holes}

\begin{figure*}[ht!]
\plottwo{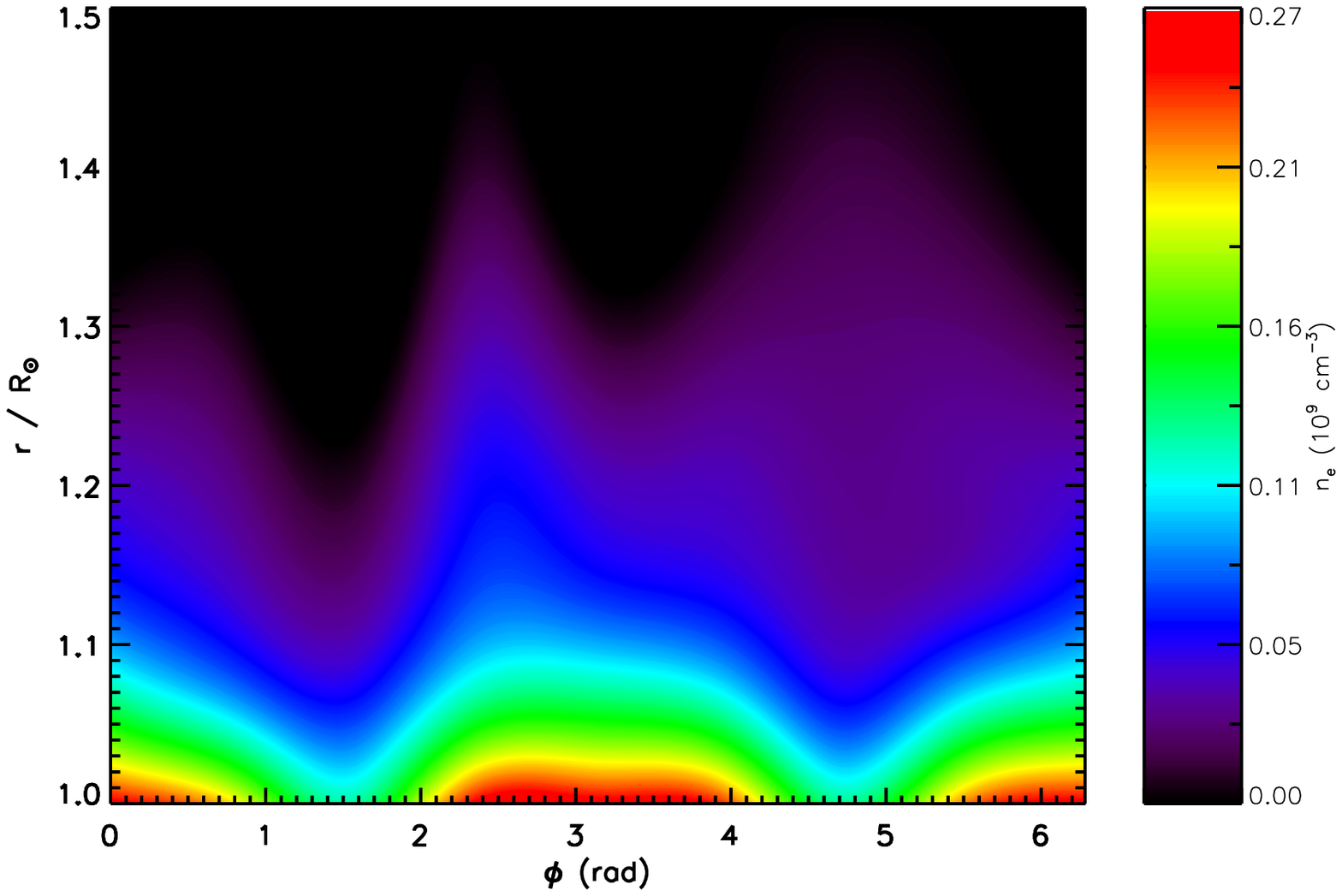}{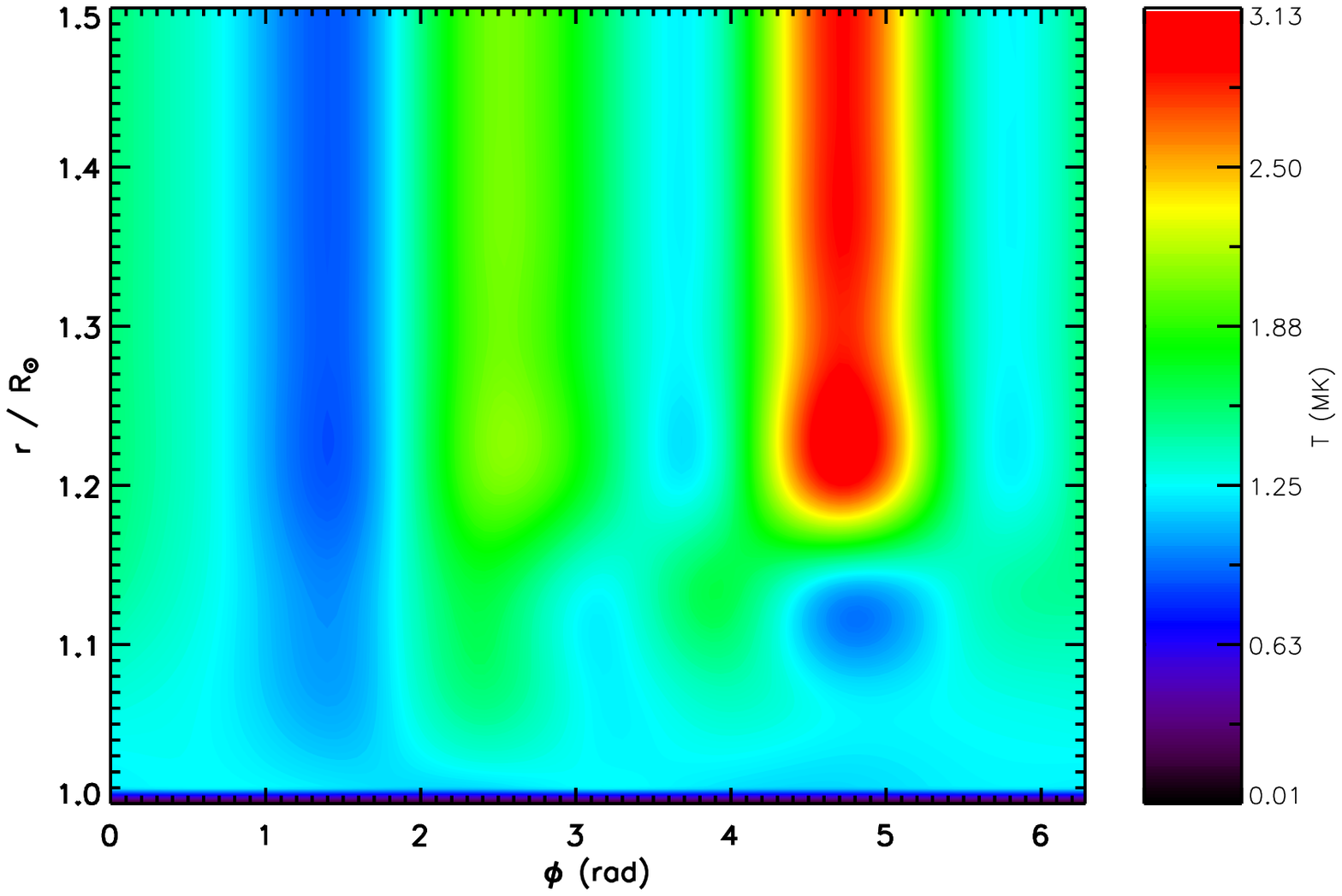}
\plottwo{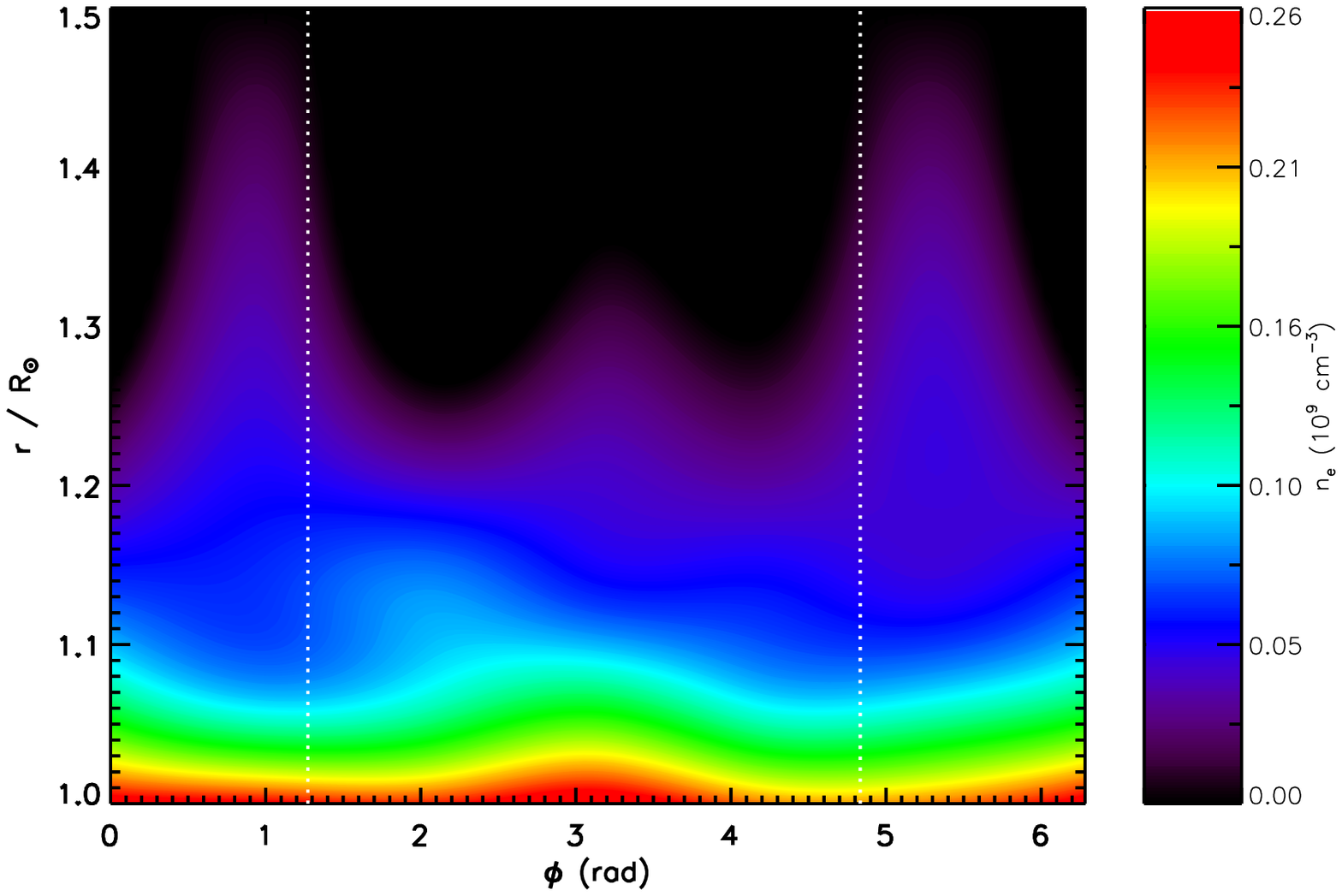}{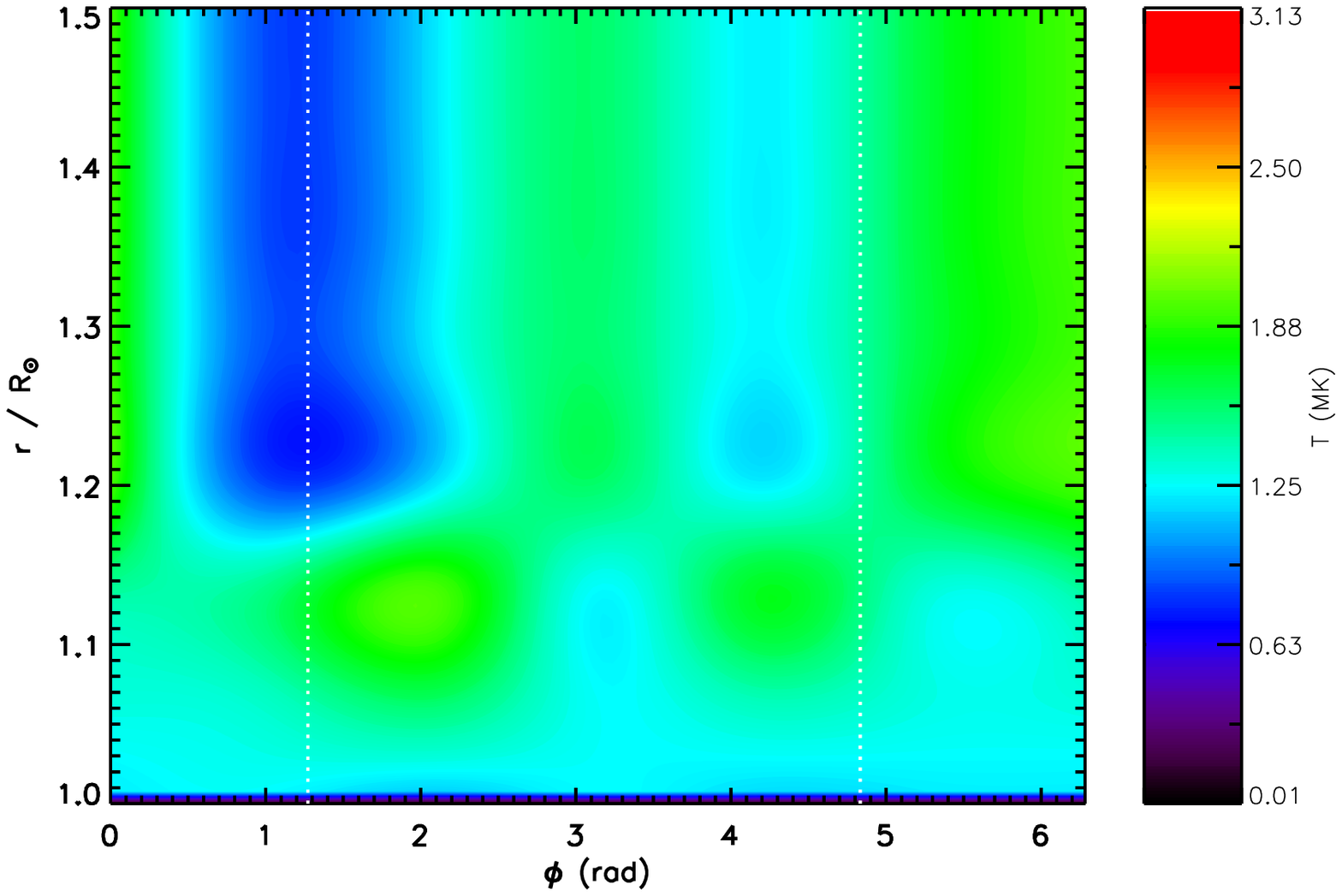}
\caption{Maps of the background coronal density and temperature for a model without coronal holes (\textit{top} panels) and a model with two coronal holes (\textit{bottom} panels) which are located at $\phi \approx 1.3$ and $4.8$~rad (vertical dotted lines).}
\label{fig:model_maps}
\end{figure*}

\begin{figure}[ht!]
\plotone{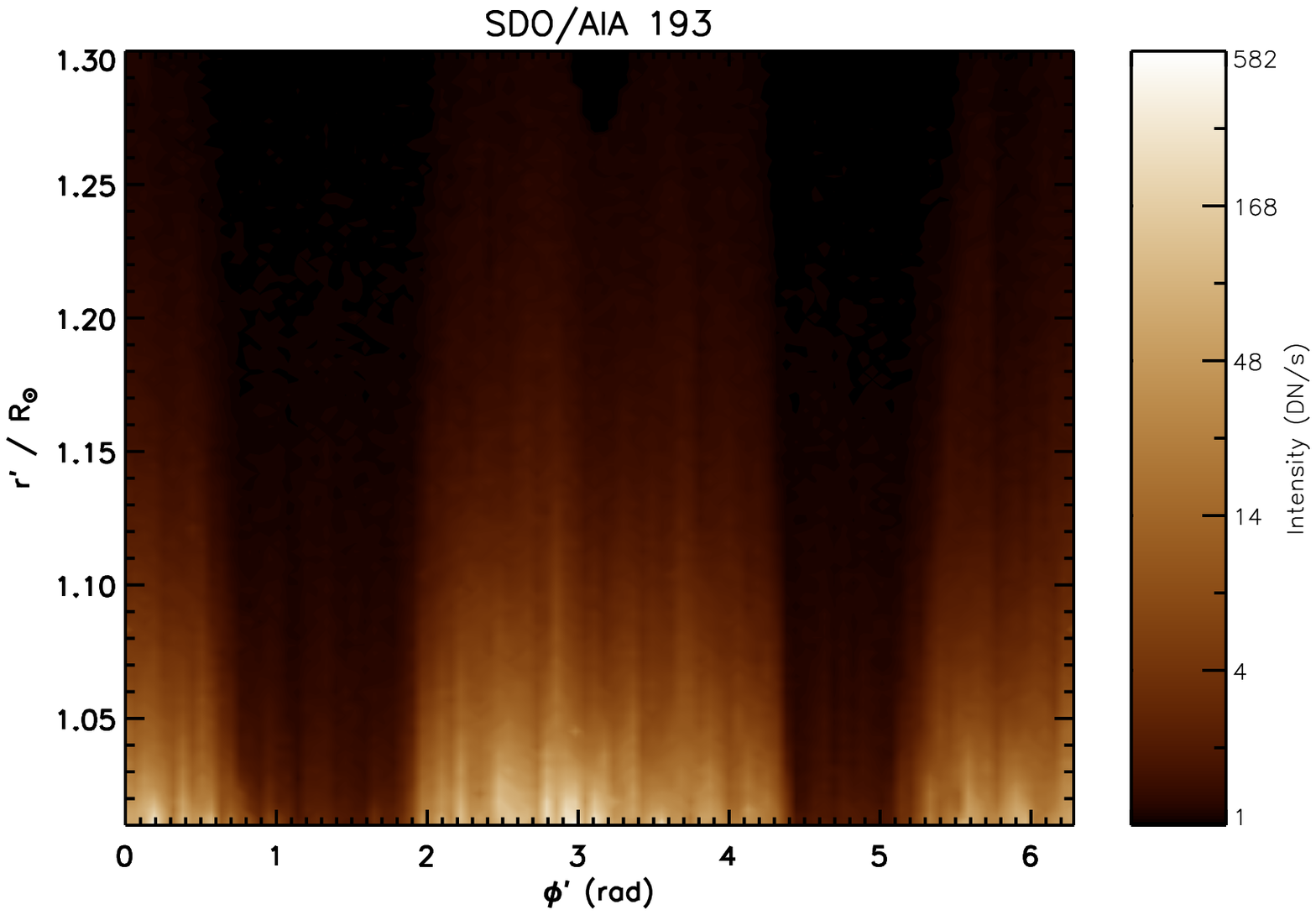}
\plotone{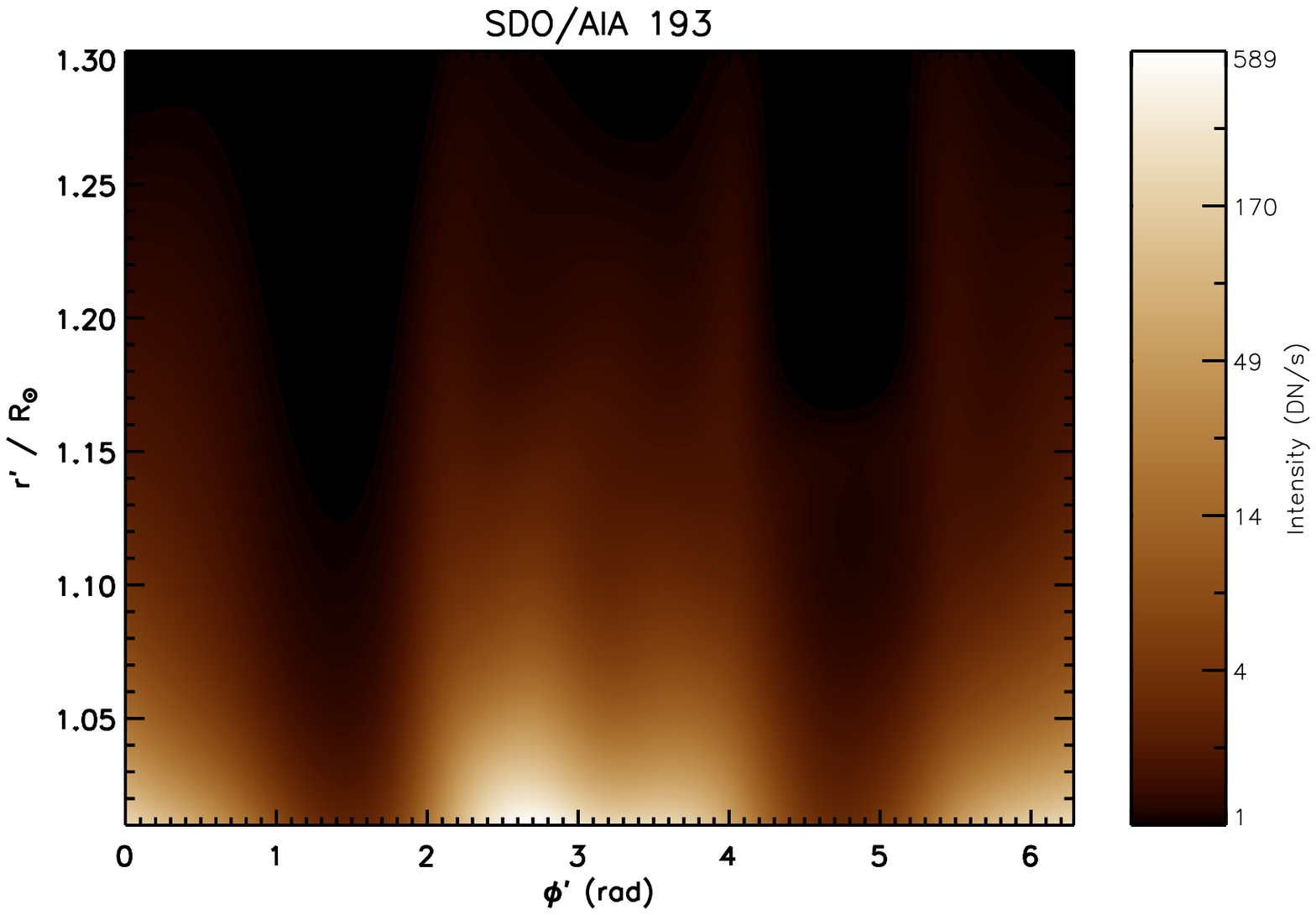}
\plotone{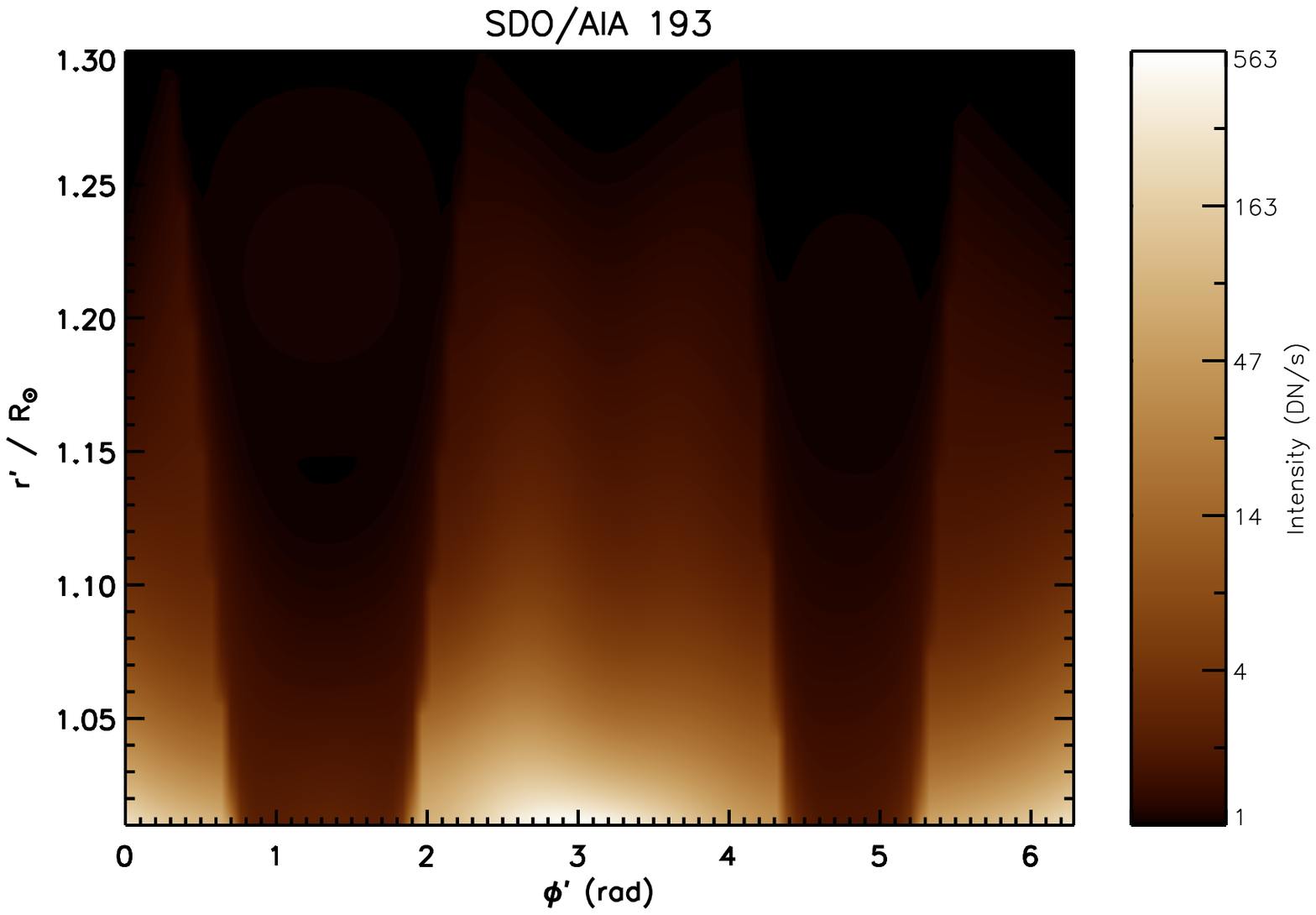}
\caption{Maps of $193$~{\AA} emission for our data (\textit{top}) and the models without (\textit{middle}) and with (\textit{bottom}) coronal holes.}
\label{fig:hole_maps}
\end{figure}

In this section we extend our model of the quiet corona to include coronal holes.
Our data (Figure~\ref{fig:maps}) exhibits two well-defined polar coronal holes and an otherwise quiet (above the limb) corona.
Modeling the size of these structures requires analysis of 2D intensity maps rather than the 1D intensity profiles considered in Section~\ref{sect:quiet}.
First, we may simply extend our previous method to apply it to 2D intensity maps by calculating the radial density and temperature profiles independently for each value of $\phi$.
Alternatively, we can choose to describe the corona with a lower number of density and temperature profiles at various positions in $\phi$, with the values of density and temperature values at locations in between being determined by interpolation. In this way every position in $\left( r',\phi' \right)$ is forward modeled and compared to data but the number of model parameters can be significantly reduced.
The separation between the model density and temperature profiles in the $\phi$-direction defines the spatial scale on which the model can resolve features. In the case of a perfectly quiet corona only a single pair of profiles would be required since the atmosphere would be spherically symmetric.
The top panels of Figure~\ref{fig:model_maps} shows maps of density and temperature emission created by this method, using $8$ pairs of density and temperature profiles across the entire corona $\phi = \left[ 0,2\pi \right]$, and spline interpolation of densities and temperatures with a periodic boundary in the $\phi$-direction.
The corresponding forward modeled emission is shown in the middle panel of Figure~\ref{fig:hole_maps} (with the observational data in the top panel).
The influence of the two polar coronal holes is evident, being regions of lower density and temperature than the surrounding corona \citep[e.g.][]{1972ApJ...176..511M,1975SoPh...40..351W}.
The emissivity from the plasma comprising a coronal hole is typically significantly less than the surrounding plasma, and so by taking the place of plasma which would otherwise produce far more emission it leads to a reduction in the integrated LOS intensity.
A review of coronal holes is provided by \citet{2009LRSP....6....3C}.
The lower density of coronal holes causes them to act as fast magnetoacoustic anti-waveguides \citep[e.g.][]{1995JGR...10023413O} and externally excited fast waves are observed to reflect away from them \citep[e.g.][]{1998GeoRL..25.2465T,Gopalswamy_2009}.
Nonetheless MHD wave activity may be observed in \cite[e.g.][]{2011SSRv..158..267B} or along the outer edge \citep{2014A&A...568A..20P} of coronal holes and so accurate estimation of their properties has potential seismological application.

In the case of exceptionally large coronal holes, our method in its current form may allow good estimates of the density and temperature profiles for observations near their center.
This would be the case in which the emission from the background corona is negligible and so emission from a single plasma is measured along the LOS, as assumed by our quiet corona model.
More generally the observed emission will contain significant contributions from both the coronal hole and the background and so it is necessary to model each.
Even in the case of very large holes, the contribution from the background corona becomes increasingly significant towards its boundary where the LOS depth of the hole decreases.

We now further extend our model to take the presence of coronal holes into account, in particular their finite size and super-radial expansion which do not satisfy the assumption of azimuthal symmetry in our quiet corona model.
We use a simple model for a coronal hole as an expanding flux tube.
Our data features two polar coronal holes and so we include two additional structures in addition to the background corona.
Each hole is described as having a circular cross-section centred on some position in $\phi$.
The location and size of the coronal holes are fitted parameters of the model along with the density and temperature profiles.
In this analysis, the position along our LOS is not considered and for simplicity the axis of the coronal hole is assumed to be perpendicular to our LOS.
The radius of the hole varies with height to describe the (super-radial) expansion of the holes.
The maps of coronal density and temperature for this new model are shown in the bottom panels of Figure~\ref{fig:model_maps}, and the forward modeled intensity in the bottom panel of Figure~\ref{fig:hole_maps}.
For a fairer comparison with the model without holes, we use $6$ pairs of radial profiles for the background density and temperature, giving the same total of $8$ pairs when including those for the holes.
The density and temperature profiles for this model show less variation near the locations of the holes (indicated by the vertical dotted lines).
Some variation is still present, presumably owing to the limitations of the simple model we use for the holes rather than being actual variations in the background that happen to coincide with them.
The large temperature variations found for the model without holes is also no longer present, particularly the high seemingly unphysical large temperature near the hole at $ \approx 4.8$~rad. We may consider these to have been a result of the discrepancy between the assumption of azimuthal symmetry and the presence of a coronal hole, particularly for large heights where low EUV intensities have larger uncertainties.
The model including the coronal holes more accurately reproduces the EUV appearance of these structures in Figure~\ref{fig:hole_maps}, particularly with regard to the sharpness of the boundaries of the holes.
Our coronal hole model also allows the super-radial expansion of the hole with height to be reproduced, which cannot be described by our model only describing a quiet corona.

\begin{figure*}[ht!]
\plottwo{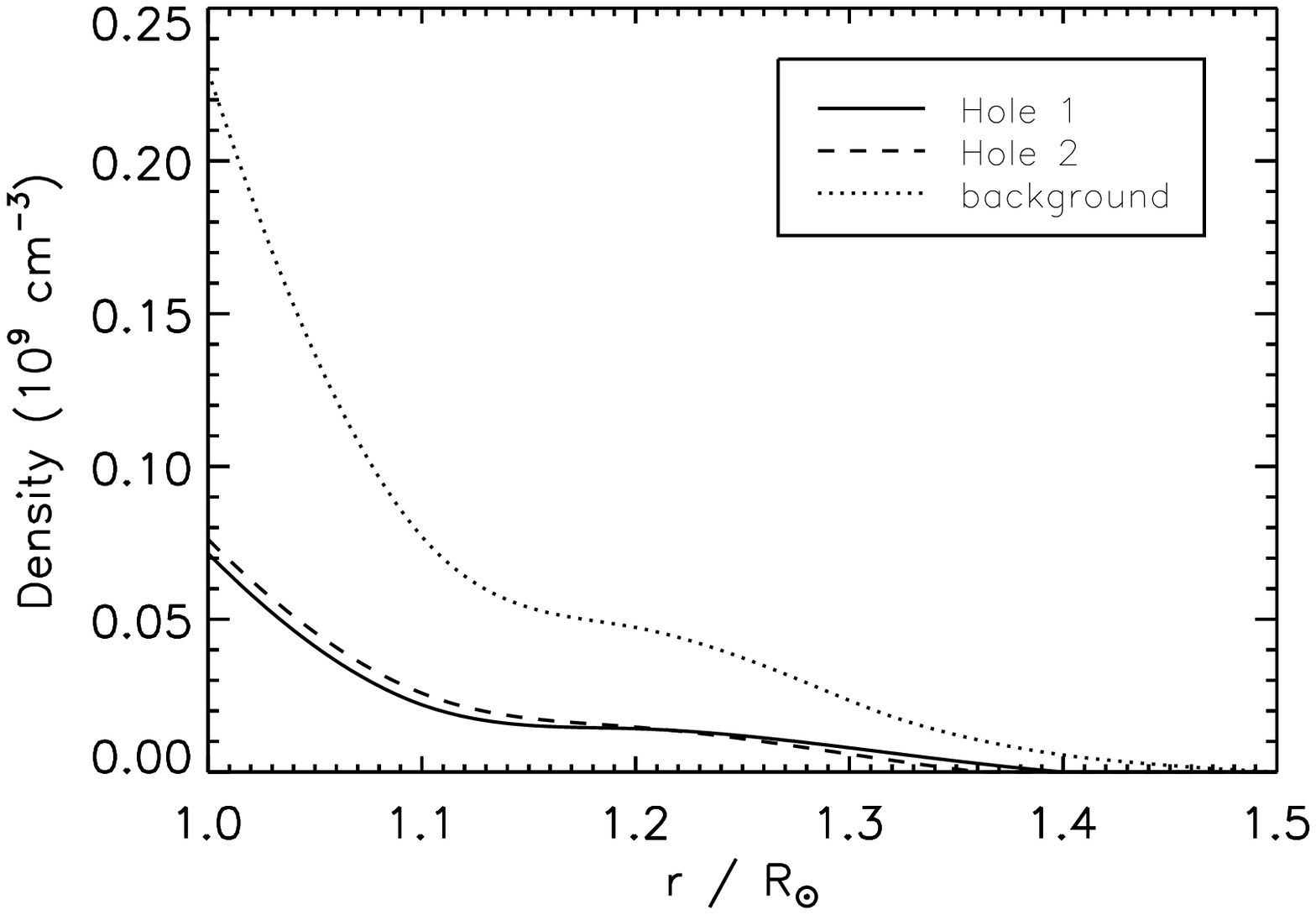}{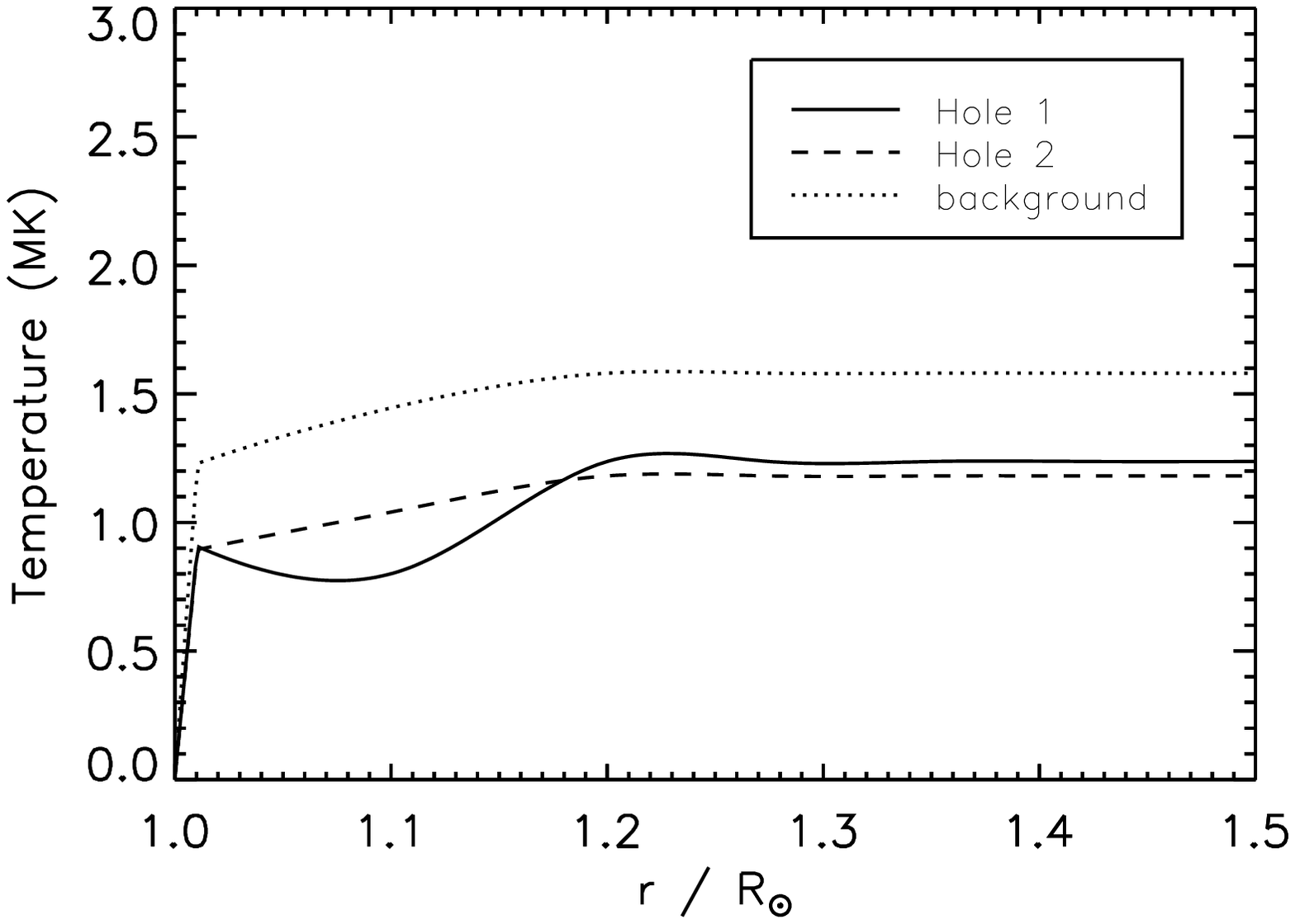}
\caption{Density and temperature profiles for the coronal holes at $\phi \approx 1.3$~rad (solid) and $\phi \approx 4.8$~rad (dashed).
The dotted lines denote the average profiles for the background plasma.}
\label{fig:holes_plots}
\end{figure*}

Figure~\ref{fig:holes_plots} shows the density and temperature profiles for each of the modeled coronal holes.
The dotted lines show the corresponding mean profiles (averaged over $\phi$) for the background plasma.
These profiles show the expected behaviour of the holes having lower density and temperature than the surrounding plasma.
We can estimate that the two holes have density contrast ratios of approximately $0.30$ and $0.33$, and temperature contrast ratios of $0.65$ and $0.73$, compared to the respective values for background corona.
For our model that does not include coronal holes, the fitted plasma densities typically fall between the values of the holes and background, representing an averaged value due to the effect of LOS integration.

Our polar coronal hole densities of approximately $10^{8}$~cm$^{-1}$ and less are consistent with those found by spectral diagnostics by \citet{1997ApJ...482L.109D} who used \ion{Si}{8} and \ion{S}{10} line ratios
\citep[see also recent review of spectral diagnostics by][]{2018LRSP...15....5D}.
The coronal hole temperatures above $1$~MK we find are typically slightly higher than those founds by other authors \citep[e.g.][]{1998A&A...336L..90D}.
On the other hand, other estimates of temperatures less than $1$~MK \citep{1998ApJ...500.1023W} have found to be revised to just above $1$~MK by improvement of the atomic calculations \citep{2008A&A...487.1203D}.
\citet{2013ApJ...776...78H} found temperatures in a polar coronal hole to be $1$--$2$~MK.
A benefit of our method is that it naturally distinguishes the LOS emission from the holes from that from the hotter surrounding corona and so potentially allows more accurate diagnostics.
However, this depends on the systematic error due to the simplicity of the forward model used, for example in this paper we consider holes to be monolithic whereas \citep{2006A&A...455..697W} use spatial information to distinguish between plume and inter-plume regions.

\section{Conclusions}
\label{sect:conclusions}

In this paper we have demonstrated a method of estimating the density and temperature profiles of the solar corona using only optically thin EUV emission.
Our model uses spatial profiles of EUV intensity to infer the radial density and temperature profiles of the solar corona.
The assumption of a radial dependence for density and temperature corresponds to the quiet corona, i.e. without magnetic structures.
However, this may be used to model the background intensity with additional terms included to describe particular EUV structures.
We demonstrated this with a simple model for coronal holes as expanding flux tubes and applied it to describe coronal images featuring polar coronal holes.

The methods demonstrated in this paper make use of spatial information, i.e. one-dimensional data for the quiet corona and two-dimensional information when including structures with finite size such as coronal holes.
In contrast, DEM inversion methods are typically zero-dimensional and do not take any spatial dependence into account.
DEM analysis therefore considers an underdetermined system and so requires special inversion techniques \citep[e.g.][]{2012A&A...539A.146H,2015ApJ...807..143C}.
DEM analysis can be readily applied to EUV data without the need for particular models, although the interpretation of the results can be more difficult.
Particular caution is required for interpreting temperature histograms since the DEM method can overestimate their broadness, as demonstrated previously by \citet{2018A&A...620A..65V} and in Figure~\ref{fig:dem} of this paper.
The interpretation of the results of our forward modeling is typically simpler, but also requires models of sufficient detail to be developed in the first place.
A benefit of our problem being overdetermined is the ability to exclude one or more channels from the analysis. Using fewer channels requires less data to be forward modeled and so decreases the runtime (though it remains significantly longer than DEM inversions).
It is also useful if the data is unavailable for some reason, or for checking the robustness of results; for example \citet{2015A&A...582A..56D}
caution users when using empirically adjusted $094$ and $131$~{\AA} emission to correct for missing lines \citep{2014SoPh..289.2377B}.

The density and temperature of the solar corona may also be determined using white light measurements \citep[e.g.][]{1995ApJ...447L.139F,1999ApJ...510L..63E}, multifrequency radio emission \citep[e.g][]{2015A&A...583A.101M}, or more recently the using the Interface Region Imaging Spectrograph \citep[IRIS;][]{2018ApJ...864...21K}.
\citet{2018ApJ...862...18D} recently investigated on the existence of fine structure in the corona extending to several times the solar radius and so there is potential to apply models to become significantly more detailed than the demonstrations in this paper.
The simultaneous use of other observational data to reconstruct the coronal density and temperature profiles would also be useful since our method using SDO/AIA EUV emission is most sensitive to low heights $r \lesssim 1.2 R_{\odot}$.
Since our method uses EUV data alone to constrain the coronal density and temperature profiles, with no particular physical assumptions, it could readily be extended to include other sources of information.
For example, \citet{2018ApJ...860...31P} apply a method to determine the transverse structuring of coronal loops by the simultaneous use of EUV intensity profiles and seismological information from the damping profile of kink oscillations \citep[e.g.][]{2013A&A...551A..40P,2016A&A...589A.136P,10.3389/fspas.2019.00022}.
Additional information may also be provided by magnetic extrapolation \citep[e.g.][]{2013ApJ...767...16V,2018ApJ...852..137R} or by simultaneously forward modeling the observables of another instrument, e.g. radio emission observed by LOFAR \citep{2018A&A...614A..54V}.

\begin{acknowledgements}
DJP and TVD were supported by the GOA-2015-014 (KU Leuven) and the European Research Council (ERC) under the European Union's Horizon 2020 research and innovation programme (grant agreement No 724326).
The data is used courtesy of the SDO/AIA team.
The authors thank A. Dompas, W. Jacobs, L. Rollier, and P. Vanmechelen for their contributions.
CHIANTI is a collaborative project involving George Mason University, the University of Michigan (USA), University of Cambridge (UK) and NASA Goddard Space Flight Center (USA).
\end{acknowledgements}

\bibliographystyle{aa}
\bibliography{pascoe.bib,solar.bib}

\end{document}